\documentclass[floatfix, prl, twocolumn, showpacs, showkeys, preprintnumbers, nofootinbib, superscriptaddress]{revtex4-1}
\usepackage[utf8]{inputenc}
\usepackage[sort&compress]{natbib}
\usepackage{ulem}
\usepackage{bm}
\usepackage{overpic}
\usepackage{times}
\usepackage{amssymb,amsbsy,amsmath,amsfonts}
\usepackage{graphicx}
\usepackage{float}
\usepackage{color}
\usepackage{morefloats}
\usepackage{rotating}
\usepackage{srcltx}
\usepackage{slashed}
\usepackage{subfigure}
\usepackage{multirow}
\usepackage{verbatim}
\usepackage{hyperref}
\usepackage{tabularx}

\DeclareUnicodeCharacter{2212}{-}
\begin{document}

\title{Accurate relativistic chiral nucleon-nucleon interaction up to NNLO}

\author{Jun-Xu Lu}
\affiliation{School of Space and Environment, Beihang University, Beijing 102206, China}
\affiliation{School of Physics, Beihang University, Beijing 102206, China}

\author{Chun-Xuan Wang}
\affiliation{School of Physics, Beihang University, Beijing 102206, China}

\author{Yang Xiao}
\affiliation{School of Physics, Beihang University, Beijing 102206, China}
\affiliation{Université  Paris-Saclay, CNRS/IN2P3, IJCLab, 91405 Orsay, France}

\author{Li-Sheng~Geng}
\email[]{Corresponding author: lisheng.geng@buaa.edu.cn}
\affiliation{School of Physics, Beihang University, Beijing 102206, China}
\affiliation{Beijing Key Laboratory of Advanced Nuclear Materials and Physics, Beihang University, Beijing 102206, China}
\affiliation{School of Physics and Microelectronics, Zhengzhou University, Zhengzhou, Henan 450001, China }

\author{Jie Meng}
\affiliation{State Key Laboratory of Nuclear Physics and Technology, School of Physics, Peking University, Beijing 100871, China}
\author{Peter Ring}
\affiliation{Physik Department, Technische Universit\"at M\"unchen, D-85747 Garching, Germany}
\begin{abstract}
We construct a  relativistic chiral nucleon-nucleon interaction up to the next-to-next-to-leading order in covariant baryon chiral perturbation theory. We show that  a  good description of the $np$ phase shifts up to $T_\mathrm{lab}=200$ MeV and even higher can be achieved with a $\tilde{\chi}^2/\mathrm{d.o.f.}$ less than 1. Both the next-to-leading order results  and the next-to-next-to-leading order results describe the phase shifts equally well up to $T_\mathrm{lab}=200$ MeV, but for higher energies, the latter behaves better, showing satisfactory convergence. The relativistic chiral potential provides the most essential inputs for  relativistic ab initio studies of nuclear structure and reactions, which has been in  need for almost two decades.
\end{abstract}

\maketitle

The nucleon-nucleon ($NN$) interaction plays an essential role in our microscopic understanding of nuclear physics. Starting from the pioneering works of Weinberg~\cite{Weinberg:1990rz,Weinberg:1991um,Weinberg:1992yk}, chiral effective field theory (ChEFT) has been successfully applied to derive the $NN$ interaction. Nowadays the so-called chiral nuclear forces  have been constructed up to the fifth order~\cite{Epelbaum:2014sza,Reinert:2017usi,Entem:2017gor} and sixth order~\cite{RodriguezEntem:2020jgp} and reached the level of the most refined phenomenological forces, such as Argonne $\textrm{V}_{18}$~\cite{Wiringa:1994wb} and CD-Bonn~\cite{Machleidt:2000ge}, and  have become the de facto standard in ab initio nuclear structure and reaction studies~\cite{Epelbaum:2008ga,Machleidt:2011zz,Hammer:2019poc,Drischler:2019xuo}.

Nonetheless, these forces are based on the non-relativistic heavy baryon chiral perturbation theory(ChPT) and cannot be used in relativistic many-body studies~\cite{Shen:2016bva,Shen:2017vqr,Tong:2018qwx,Tong:2019juo,Wang:2021mvg}, for which  till now only the Bonn potential~\cite{Machleidt:1987hj} has been widely used~\cite{Shen:2019dls}. In addition, there are continuing discussions on the relevance of renormalization group invariance and how the Weinberg power counting should be modified to allow for proper non-perturbative renormalization group invariance~\cite{Epelbaum:2018zli,vanKolck:2020llt}. Lorentz covariance, as one of the most fundamental requirements of Nature, may play a role here. It is particularly inspiring to note that in the one-baryon sector covariant baryon ChPT has been shown to provide new perspectives on a number of long standing puzzles, such as baryon magnetic moments~\cite{Geng:2008mf}, Compton scattering off protons~\cite{Lensky:2009uv},  pion-nucleon scattering~\cite{Alarcon:2011zs}, and baryon masses~\cite{MartinCamalich:2010fp,Ren:2012aj}. See Ref.~\cite{Geng:2013xn} for a short review.

Recently, a covariant power counting approach  similar to the extended-on-mass-shell scheme in the one-baryon sector~\cite{Gegelia:1999gf,Fuchs:2003qc} was proposed to describe the $NN$ interaction~\cite{Ren:2016jna,Xiao:2018jot}.~\footnote{We note that a modified Weinberg’s approach to the $NN$ scattering problem was
proposed in Ref.~\cite{Epelbaum:2012ua}, which employs time-ordered perturbation theory and
relies on the manifestly Lorentz-invariant effective  Lagrangian and aims at improving the UV behavior  of Weinberg’s approach~\cite{Epelbaum:2015sha,Behrendt:2016nql,Ren:2022glg}. } At leading order (LO), the covariant scheme has been successfully tested in  the $NN$ system~\cite{Ren:2016jna,Ren:2017yvw,Bai:2020yml,Bai:2021uim,Wang:2020myr}, hyperon-nucleon system~\cite{Li:2016paq,Li:2016mln,Li:2018tbt,Song:2018qqm,Liu:2020uxi,Song:2021yab}, and $\Lambda_{c} N$ system~\cite{Song:2020isu,Song:2021war}. In addition to providing already a reasonable description of the $J=0,1$ $np$ phase shifts at LO, it also shows some interesting features of  proper effective field theories. In Ref.~\cite{Ren:2017yvw}, it was shown that for the $^1S_0$ partial wave, some of the typical low energy features can be reproduced at LO, contrary to the conventional Weinberg approach. In addition, it  also shows improved renormalization group invariance, for example, in the $^3P_0$ channel~\cite{Wang:2020myr}. In Ref.~\cite{Girlanda:2018xrw}, it was shown in a hybrid phenomenological approach that  the LO relativistic three-body interaction leads to a satisfactory description of polarized $pd$ scattering data in the whole energy range below the deuteron breakup threshold, solving the longstanding $A_y$ puzzle thanks to the new terms considered in the $3N$ force.  Furthermore, in Ref.~\cite{Xiao:2020ozd}, it was shown that the relativistic effects in the perturbative two-pion-exchange (TPE) contributions do improve the description of the peripheral $NN$ scattering data compared to their non-relativistic counterparts. In Ref.~\cite{Wang:2021kos}, the same feature is found also for the non-perturbative TPE  contributions.
 
Nonetheless, for realistic studies of nuclear structure and reactions,  the relativistic chiral force has to be constructed to higher chiral orders.  Furthermore, a complete understanding of the relativistic chiral nuclear force beyond leading order is also of high relevance. For such purposes, in the present work, we construct the first accurate relativistic $NN$ interaction up to the next-to-next-to-leading order (NNLO).~\footnote{Recent studies show that the NNLO non-relativistic chiral forces can provide reliable inputs already, but the N$^3$LO forces yield smaller uncertainties~\cite{Huther:2019ont}}

\begin{figure}[h]
\centering
\includegraphics[width=0.46 \textwidth]{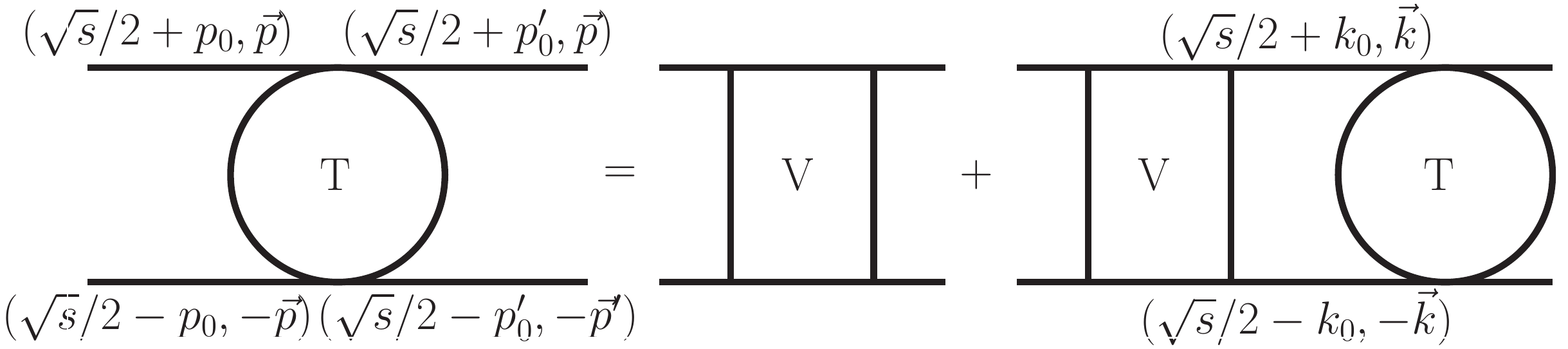}
\caption{Relativistic kinematics of nucleon-nucleon scattering.}
\label{kinematics}
\end{figure}

In order to take into account the non-perturbative nature of the $NN$ interaction, we solve the following relativistic Blankenbecler-Sugar (BbS) equation~\cite{Blankenbecler:1965gx}, 
\begin{equation}\label{BbSE}
\begin{split}
  T(\bm{p}',\bm{p},s)&=V(\bm{p}',\bm{p},s)\\
  +\int\frac{\mathrm{d}^3\bm{k}}{(2\pi)^3}&V(\bm{p}',\bm{k},s)\frac{m^2}{E_k}\frac{1}{\bm{q}_{cm}^2-\bm{k}^2-i\epsilon}T(\bm{k},\bm{p},s),
\end{split}
\end{equation}
where $|\bm{q}_{cm}|=\sqrt{s/4-m^2}$ is the nucleon momentum on the mass shell in the center of mass (c.m.) frame, and a sharp cutoff $\Lambda$ is introduced to regularize the potential and its value will be specified later. The momenta of the incoming, outgoing, and intermediate nucleons are depicted in Fig.~\ref{kinematics}, consistent with the 3D reduction of the Bethe-Salpeter equation to the BbS equation~\cite{Woloshyn:1973mce}.

Up to NNLO, the relativistic chiral potential consists of the following terms
\begin{equation}\label{NNForce}
  V=V_{\mathrm{CT}}^{\mathrm{LO}}+V_{\mathrm{CT}}^{\mathrm{NLO}}+V_{\mathrm{OPE}}+V_{\mathrm{TPE}}^{\mathrm{NLO}}+V_{\mathrm{TPE}}^{\mathrm{NNLO}}-V_{\mathrm{IOPE}},
\end{equation}
in which the first two terms refer to the LO [$\mathcal{O}(p^0)$] and NLO [$\mathcal{O}(p^2)$] contact contributions, while the next three terms denote the one-pion exchange (OPE), leading, and subleading TPE contributions. The last term represents the iterated OPE contribution.

The chiral effective Lagrangians for the nucleon-nucleon interaction in  covariant baryon ChPT have been constructed up to $\mathcal{O}(p^4)$ in Ref.~\cite{Xiao:2018jot}.  There are four contact terms at LO, 13 terms at NLO, and no contact terms at NNLO. As is argued in the Supplementary Material, the large subleading TPE contributions affect the  descriptions of some higher partial waves, especially  $^3P_2$. Relevant discussions in the non-relativistic cases can be found in Refs.~\cite{Epelbaum:2003gr,Epelbaum:2003xx,Epelbaum:2004fk}. In order to partially compensate the large subleading TPE contributions to the $^3P_2$ channel, we promote two nominal N$^3$LO contact terms to NLO. Considering that in our covariant power counting, the NLO contact terms already contain terms of $\mathcal{O}(p^4)$ as is depicted in the Supplementary Material, we promote the same terms but originally counted as of N$^3$LO to NLO for the $^3P_2$-$^3F_2$ partial waves. This is equivalent to removing part of the correlations between the higher order contact terms for $^3P_2$-$^3F_2$ and the other $J=2$ partial waves. Therefore, in the end, we have in total 19 low-energy constants (LECs) up to NNLO. We note that although this number is larger than that of the non-relativistic NNLO potential (9) but smaller than that of N$^3$LO (24)~\cite{Epelbaum:2004fk,Entem:2003ft}~\footnote{It should be noted that the two isospin-violating LECs are not included here. Furthermore, the results of Ref.~\cite{Entem:2003ft} with which we compare were obtained by setting $c_{2,3,4}$ semi-free.  }.  In our relativistic framework, the 19 LECs contribute to all the partial waves with total angular momentum $J\leq 2$, which makes it impossible to reorganize the LECs according to partial waves, different from the non-relativistic cases~\footnote{Note that in the non-relativistic framework, there is no LEC for the $^3F_2$ partial wave up to N$^3$LO.}. We refer to the Supplementary Material for the explicit expressions of the LO and NLO contact potentials.

For the treatment of non-perturbative OPE and TPE contributions, we refer to Ref.~\cite{Wang:2021kos}. In Table~\ref{Tab:constant}, we show the values of LECs needed to evaluate the OPE and TPE contributions. 

\begin{table}[htpb]
\centering
\caption{Decay constant $f_\pi$ (in units of MeV)~\cite{Machleidt:2011zz}, coupling constant $g_A$~\cite{Machleidt:2011zz}, and NLO $\pi N$ couplings (in units of GeV$^{-1}$)~\cite{Chen:2012nx} adopted for evaluating the OPE and TPE diagrams.}\label{Tab:constant}
\setlength{\tabcolsep}{2.4mm}{
\begin{tabular}{cccccc}
\hline\hline
$c_1$ & $c_2$  & $c_3$ & $c_4$ &$f_\pi$ &$g_A$ \\
\hline
$-1.39$ & $4.01$  & $-6.61$ & $3.92$ &$92.4$ &$1.29$ \\
\hline\hline
\end{tabular}}
\end{table}

\begin{table*}[htpb]
\centering
\caption{LECs (in units of $10^4$GeV$^{-2}$) for the relativistic LO, NLO, and NNLO results shown in Fig.~\ref{fig:EX-uncertainties}.}\label{Tab:LECs}
\setlength{\tabcolsep}{0.4mm}{
\begin{tabular}{c|ccccccccccccccccccc}
\hline\hline
& $O_1$ & $O_2$  & $O_3$ & $O_4$ & $O_5$ & $O_6$ & $O_7$ & $O_8$ & $O_9$ & $O_{10}$ & $O_{11}$ & $O_{12}$ & $O_{13}$ & $O_{14}$ & $O_{15}$ & $O_{16}$ & $O_{17}$ & $D_1$ & $D_2$ \\
\hline
LO&$-1.32$ & $-0.21$  & $-0.93$ & $0.31$ & &&&&&&&&&&&&&& \\
NLO&$-2.62$ & $9.45$  & $-5.42$ & $-6.05$ & $30.09$ & $9.02$ & $-9.19$ & $8.74$ & $4.74$ & $7.02$ & $3.52$ & $11.42$ & $-6.03$ & $-20.55$ & $-4.99$ & $-12.80$ & $6.30$ & $0.42$ & $0.28$ \\
NNLO&$-14.83$ & $-2.25$  & $-4.85$ & $6.24$ & $-0.82$ & $1.96$ & $-6.89$ & $7.19$ & $1.44$ & $3.50$ & $-8.10$ & $-9.38$ & $-4.33$ & $-12.89$ & $-12.26$ & $-11.69$ & $3.86$ & $-1.88$ & $-0.63$ \\
\hline\hline
\end{tabular}}
\end{table*}

\begin{table*}[htpb]
\centering
\caption{$\tilde{\chi}^2=\sum_i(\delta^i-\delta^i_{\rm{PWA93}})^2$ of different chiral forces for  partial waves up to $J\le 2$. \label{Tab:chi} }
\setlength{\tabcolsep}{1.6mm}{
\begin{tabular}{c|ccccccccccccc}
\hline\hline
&Total & $^1S_0$ & $^3P_0$ & $^1P_1$ & $^3P_1$ & $^3S_1$ & $^3D_1$ & $\epsilon_1$ &$^1D_2$ & $^3D_2$ &$^3P_2$ & $^3F_2$ & $\epsilon_2$  \\
\hline

NLO&17.02 & 1.02 & 7.04 & 0.46 & 0.33 &1.80 &1.69 &0.15 & 2.18& 1.35 & 0.95  &0.01 &0.04 \\
NNLO&16.61 & 0.18 & 0.30 & 1.07 & 1.55 &3.36 &0.26 &0.03 & 0.01& 9.56 & 0.01  &0.27 &0.01 \\
NR-N$^3$LO-Idaho&8.84 & 1.53 & 0.30 & 2.41 & 0.04 &2.33 & 1.00 & 0.02 &0.57 &0.42 & 0.17 &0.03 &0.02 \\
NR-N$^3$LO-EKM&16.08 & 13.45 & 0.29 & 0.34 & 0.06 &0.01 & 0.13 & 0.01 &0.02 &0.43 & 0.12 &1.22 &0.00 \\
\hline\hline
\end{tabular}}
\end{table*}

An important feature of ChEFT is that it allows for reliable uncertainty quantification. In the literature,  two different ways have been used to estimate the truncation uncertainties of chiral nuclear forces. One is varying the cutoff in a reasonable range, for example, from 450 MeV to 550 MeV~\cite{Machleidt:2011zz}. The other is to treat the difference between the optimal results obtained at different orders as the estimate of truncation uncertainties~\cite{Epelbaum:2014sza}.  
Recently, a  general Bayesian model has been proposed~\cite{Furnstahl:2015rha,Melendez:2017phj,Melendez:2019izc} and  applied in the latest non-relativistic studies~\cite{Epelbaum:2019zqc,Epelbaum:2019kcf,Maris:2020qne}, which we follow in the present work. This method is statistically well-established and can provide a statistical interpretation for the estimated uncertainties. For a detailed account of the implementation of this approach in the present study, see the Supplementary Material.

\begin{figure*}[htbp]
\centering
\includegraphics[width=0.35\textwidth]{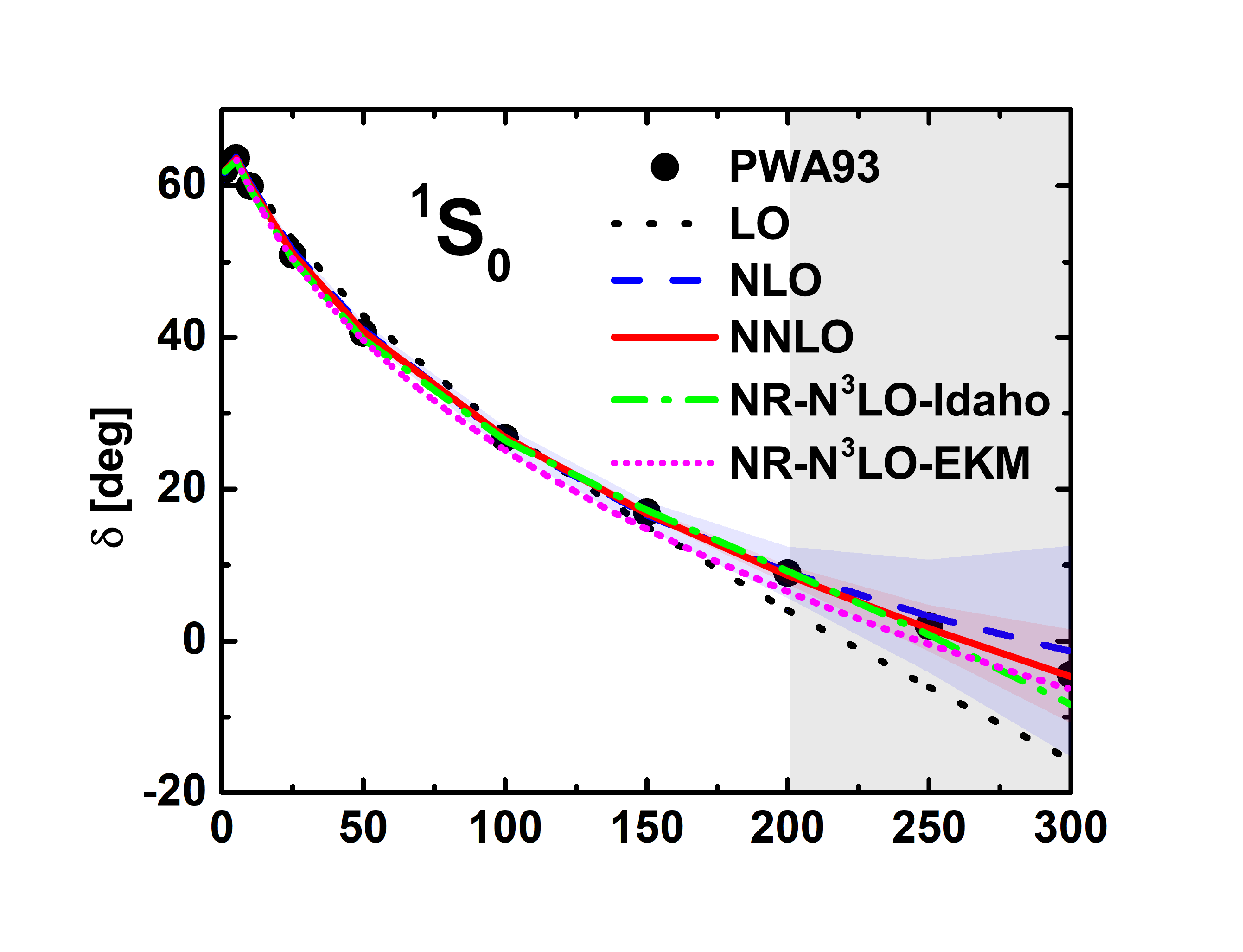}\hspace{-13mm}
\includegraphics[width=0.35\textwidth]{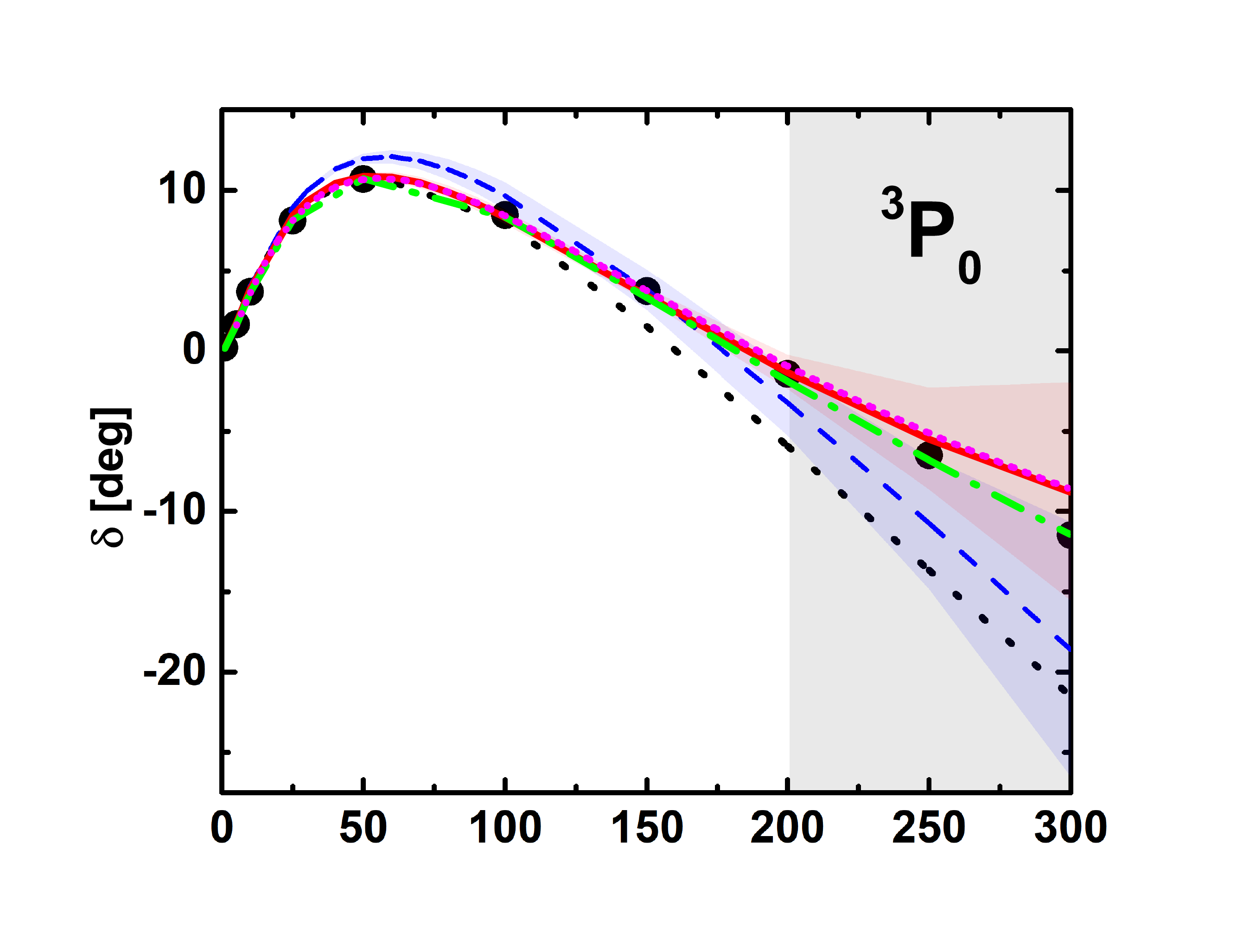}\hspace{-13mm}
\includegraphics[width=0.35\textwidth]{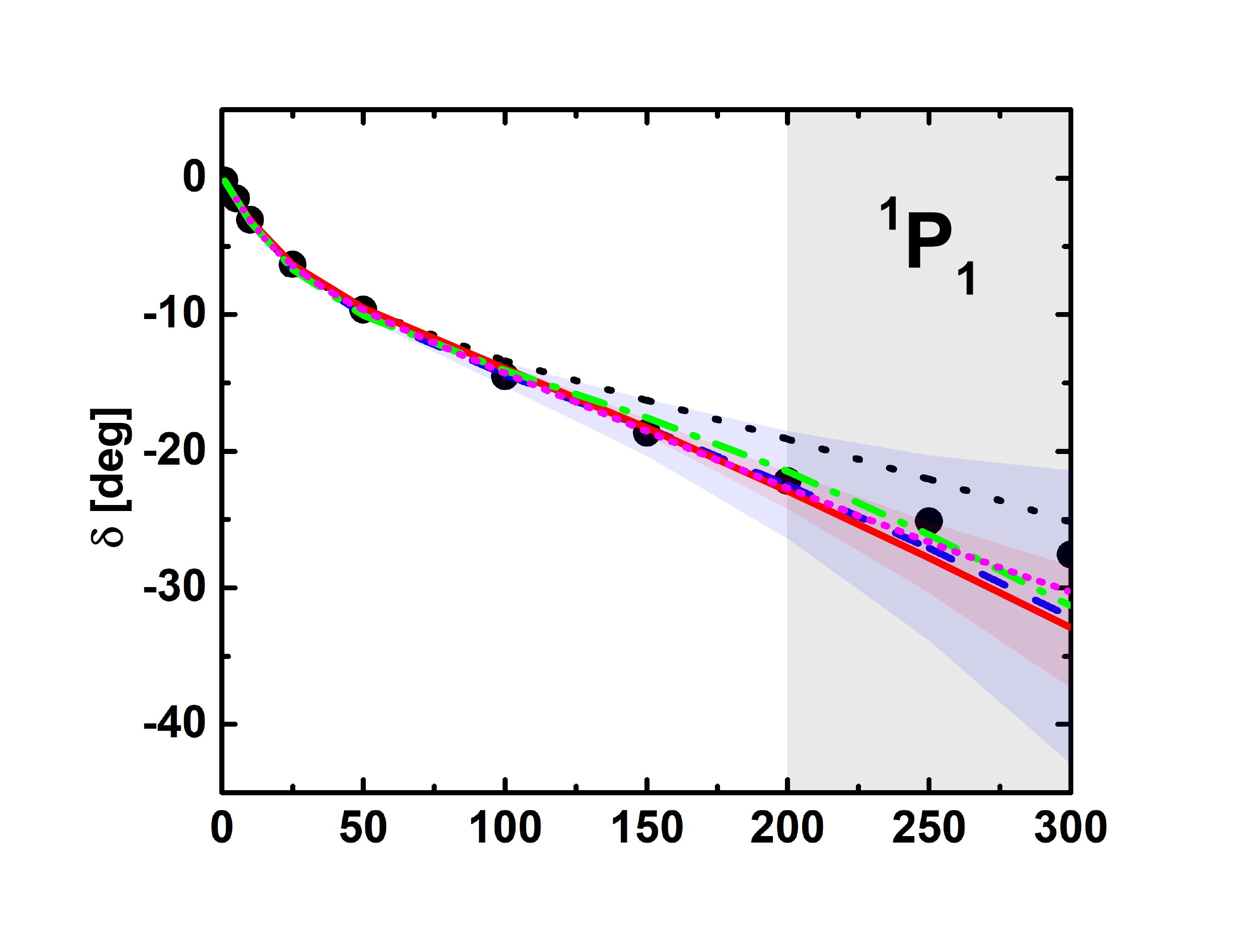}\\ \vspace{-9mm}
\includegraphics[width=0.35\textwidth]{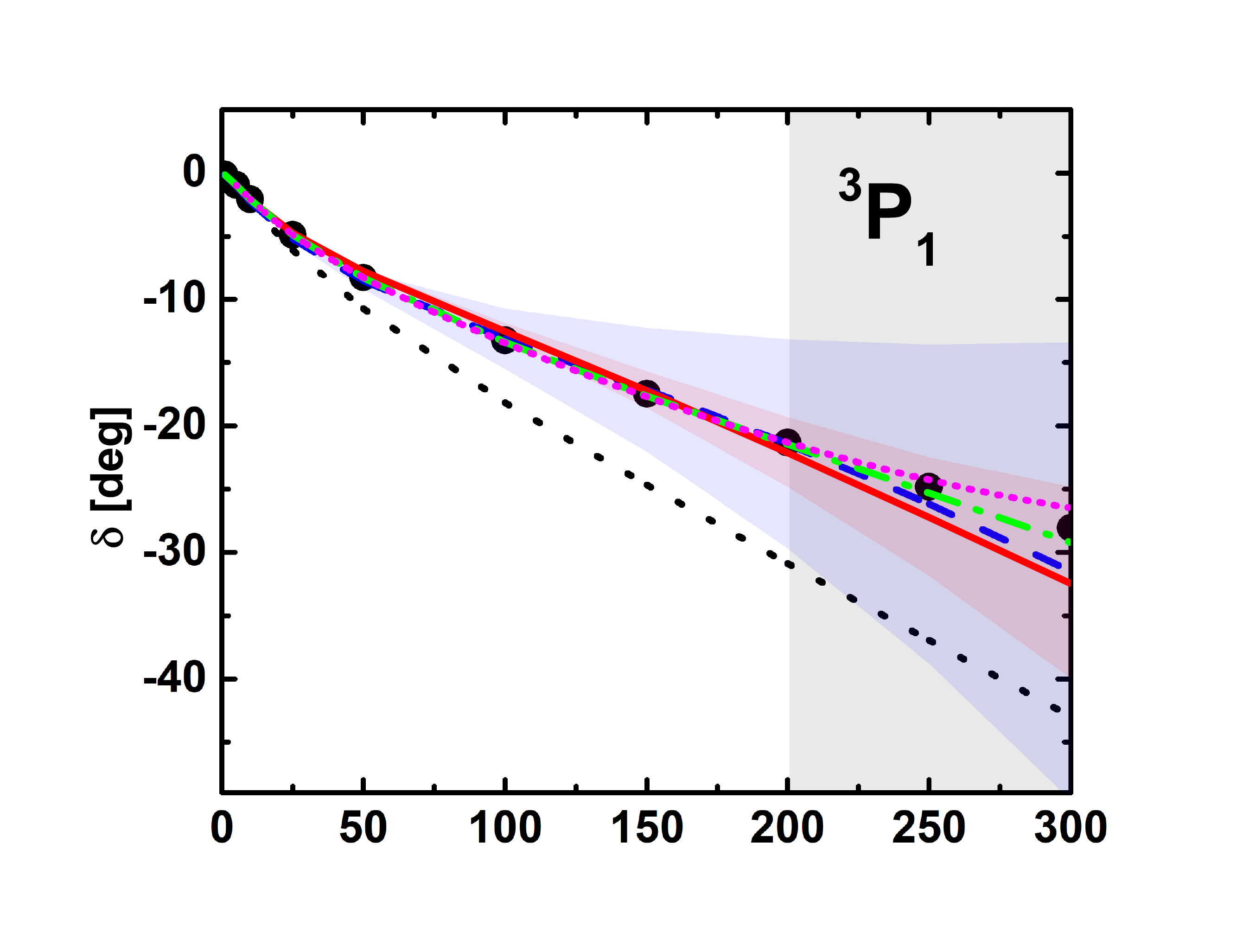}\hspace{-13mm}
\includegraphics[width=0.35\textwidth]{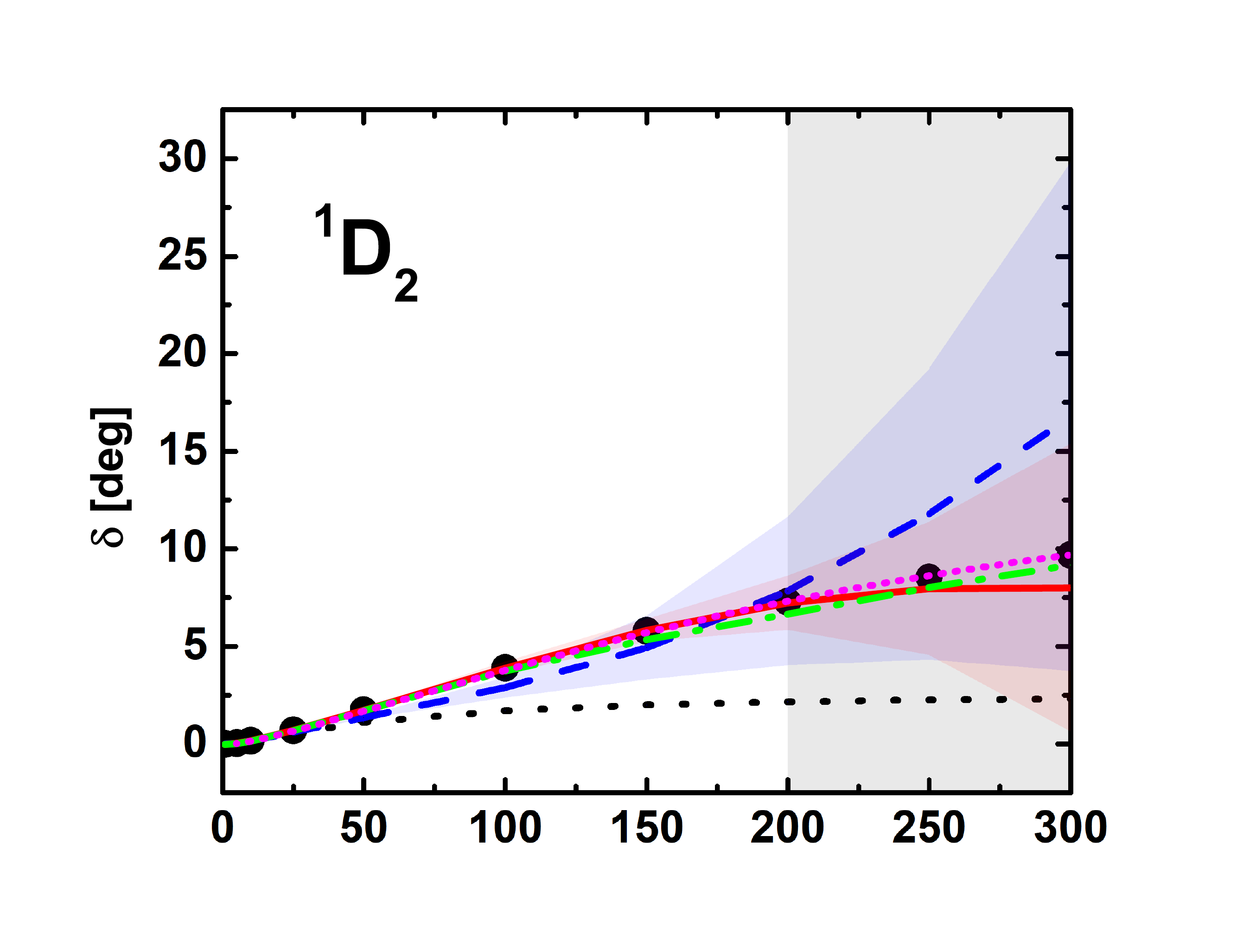}\hspace{-13mm}
\includegraphics[width=0.35\textwidth]{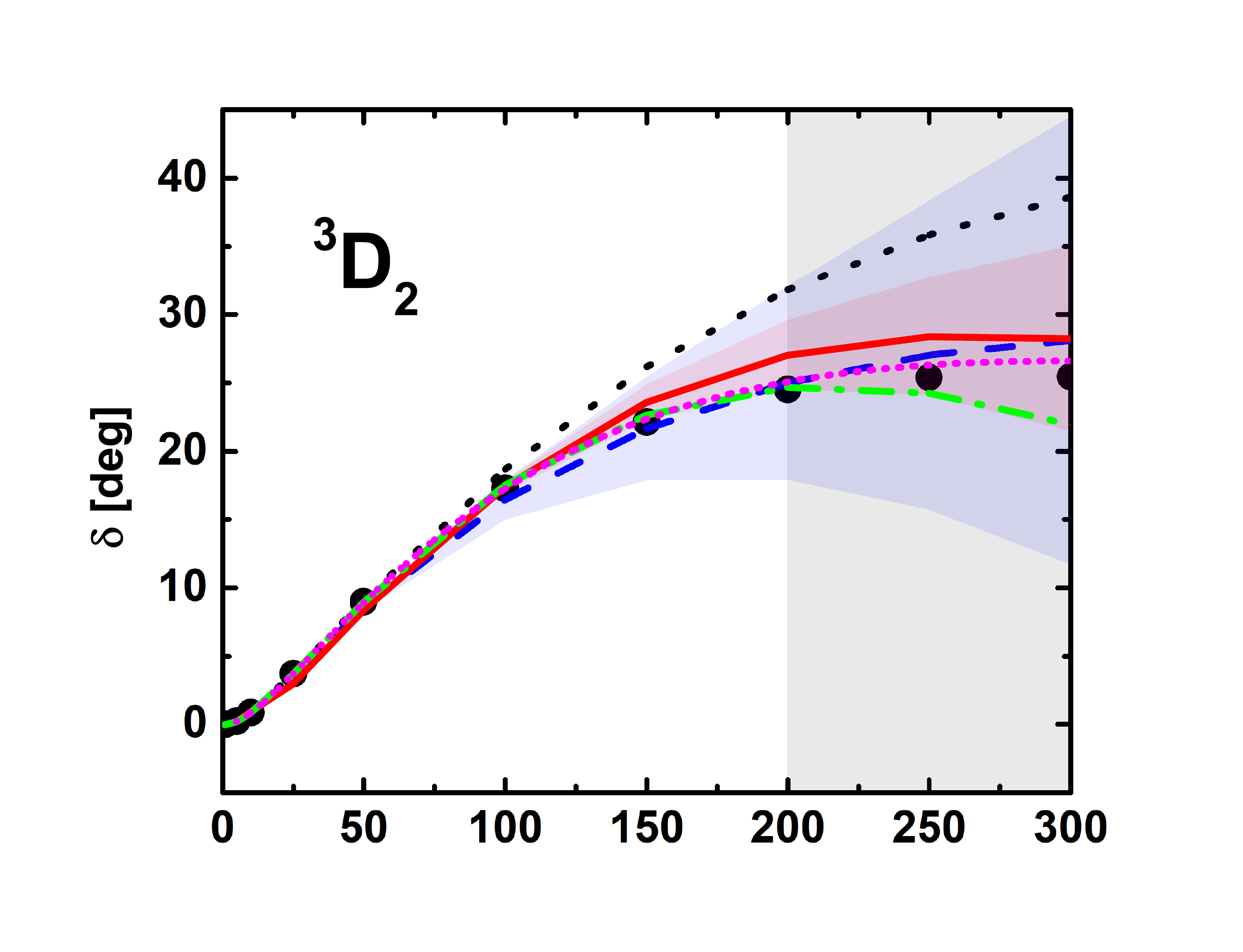}\\ \vspace{-9mm}
\includegraphics[width=0.35\textwidth]{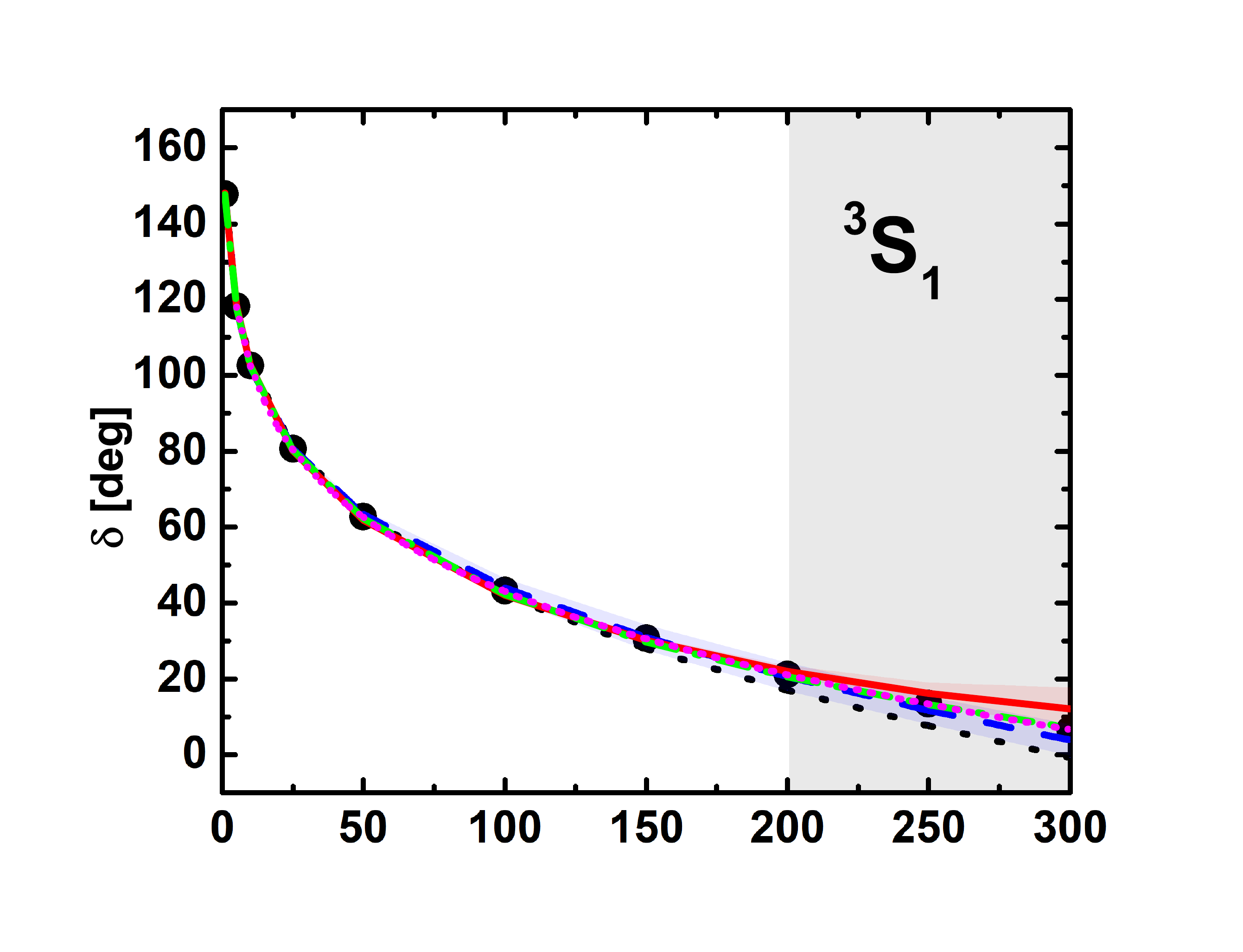}\hspace{-13mm}
\includegraphics[width=0.35\textwidth]{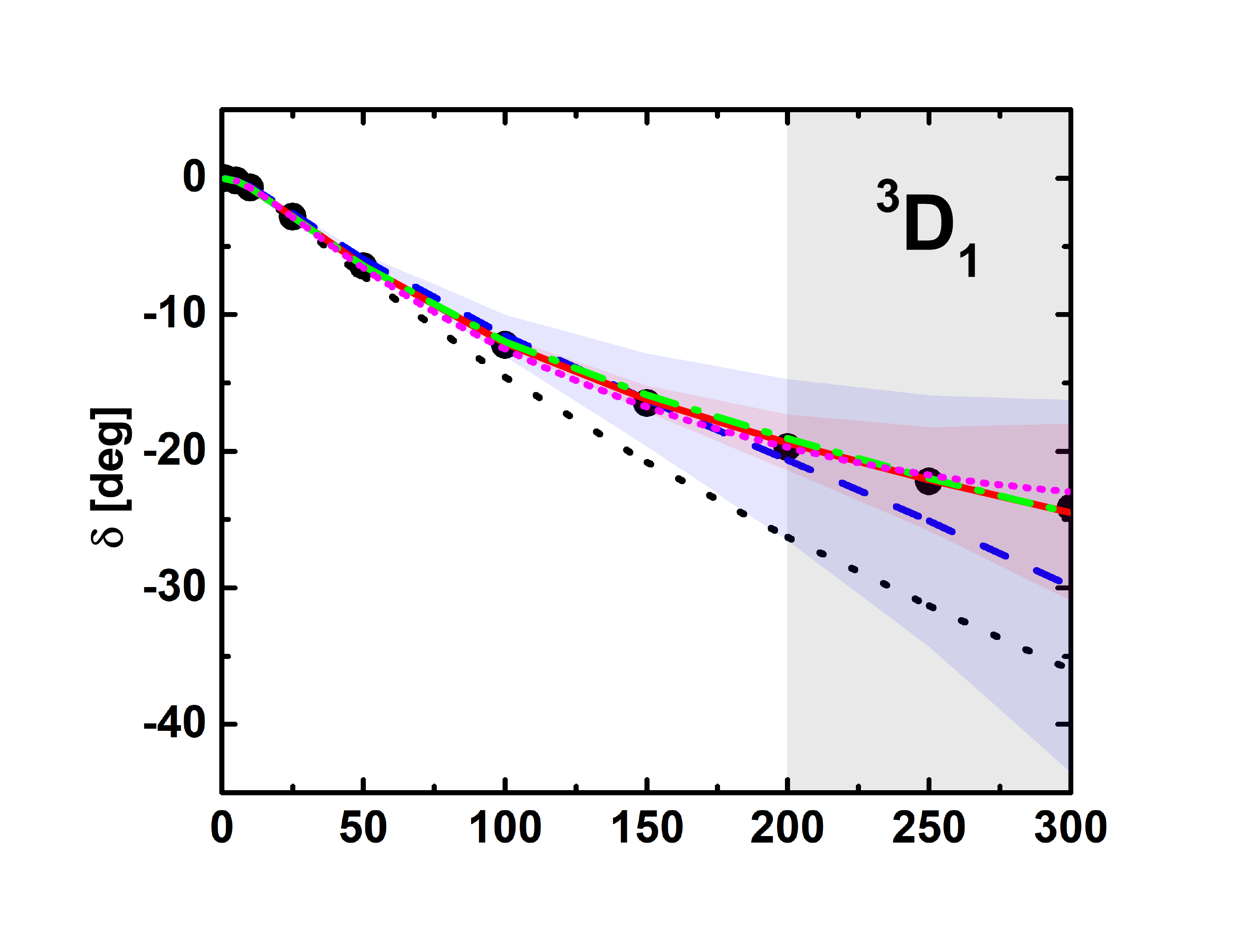}\hspace{-13mm}
\includegraphics[width=0.35\textwidth]{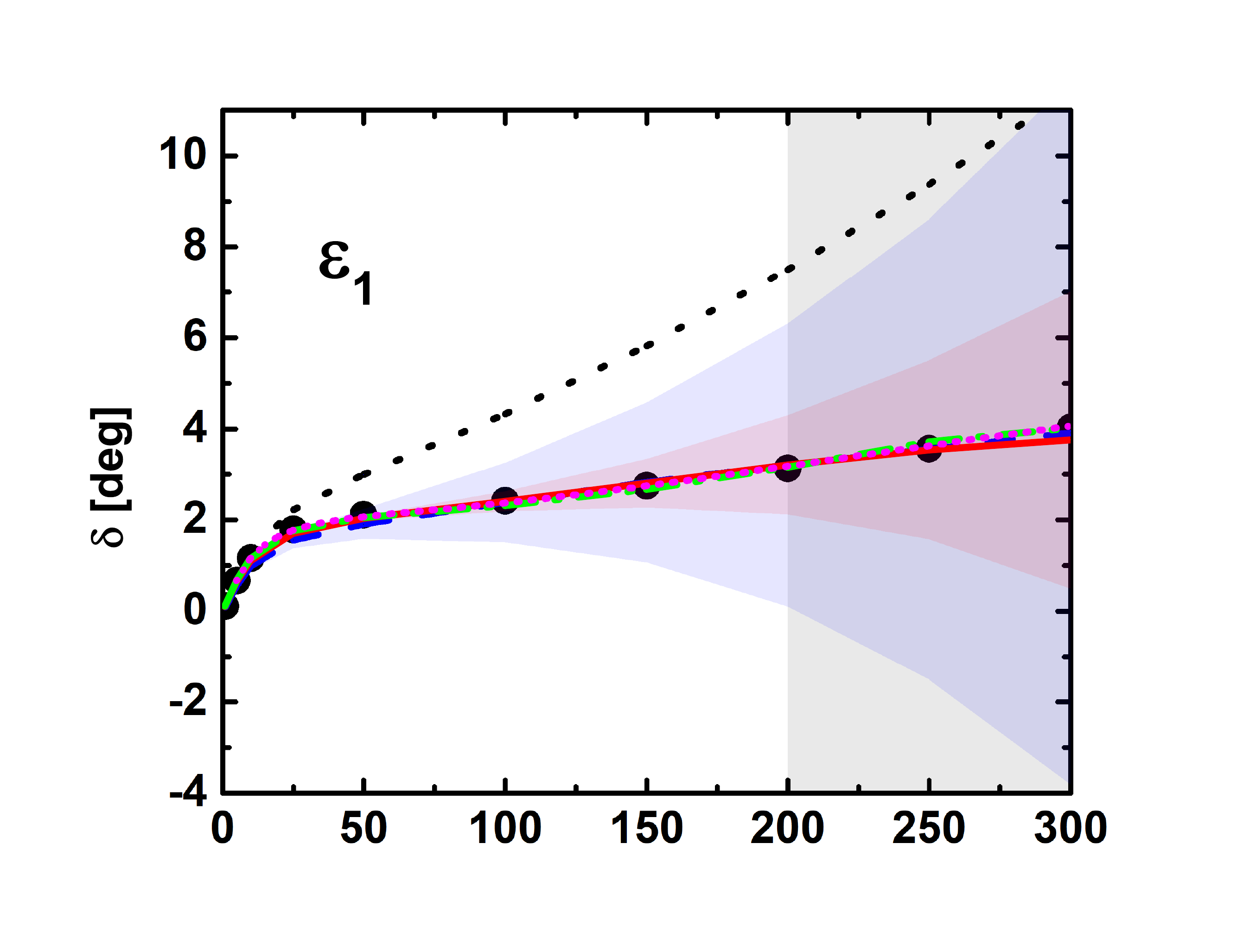}\\ \vspace{-9mm}
\includegraphics[width=0.35\textwidth]{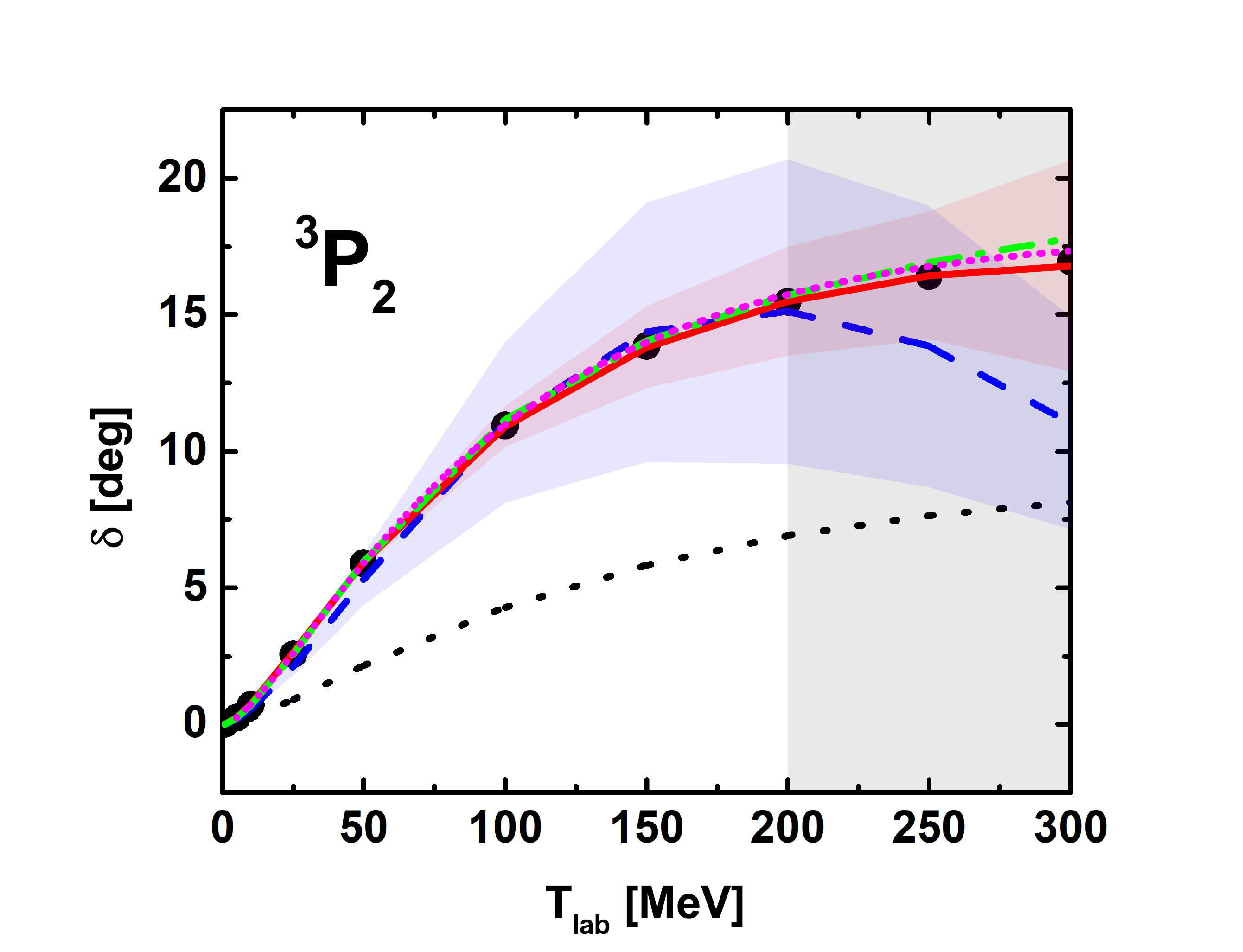}\hspace{-13mm}
\includegraphics[width=0.35\textwidth]{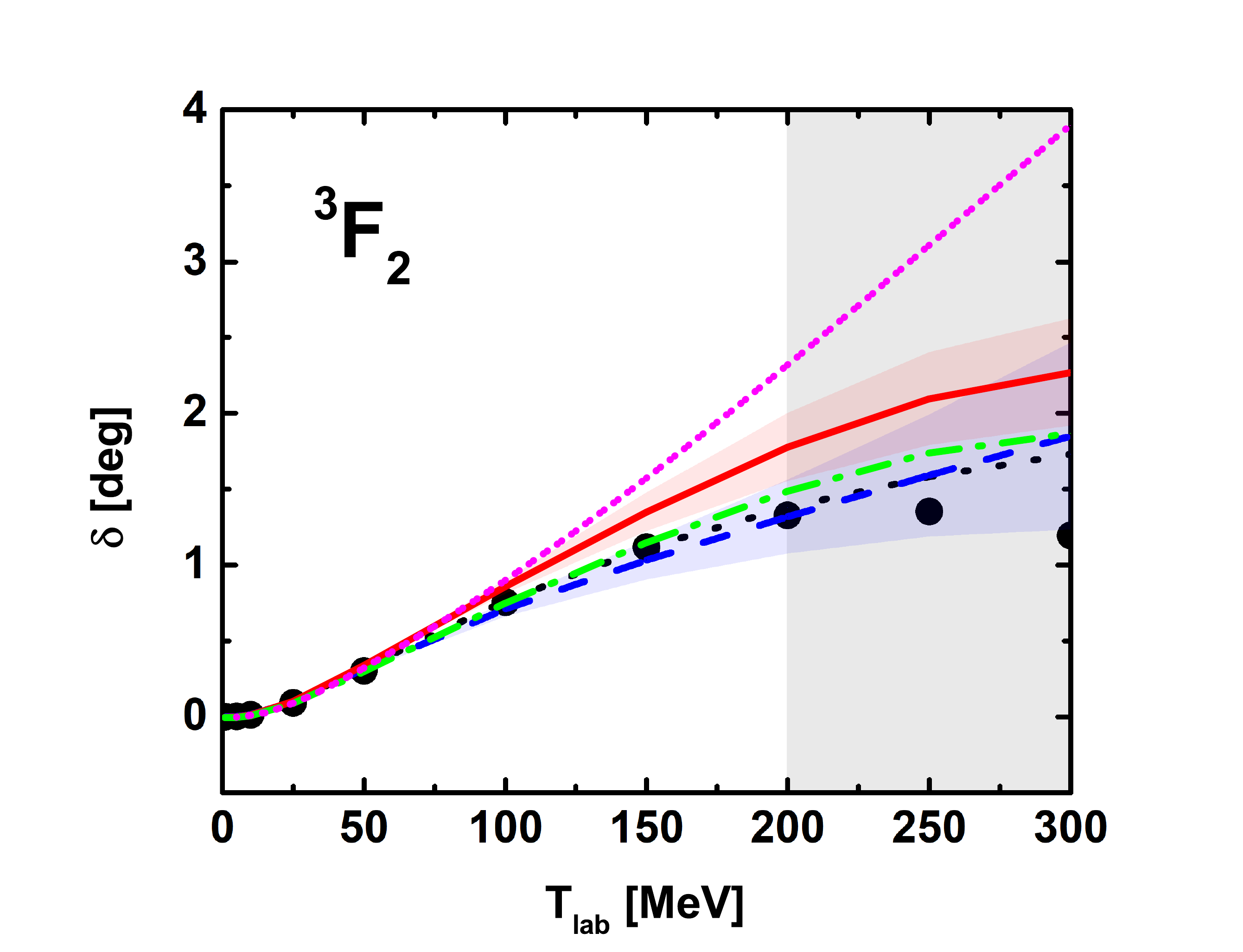}\hspace{-13mm}
\includegraphics[width=0.35\textwidth]{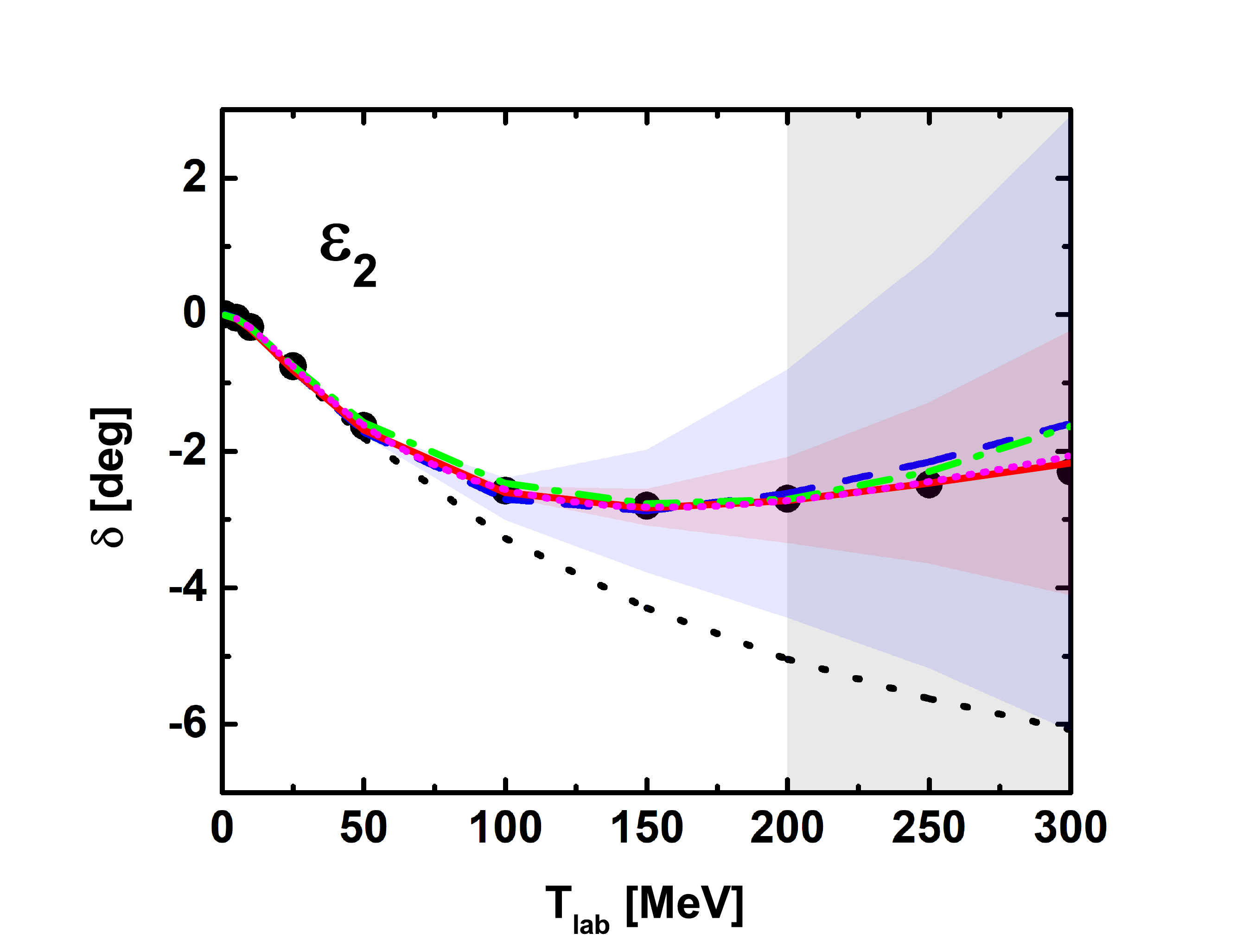}
\caption{$NN$ phase shifts for partial waves with $J\leq 2$. The red solid lines denote the relativistic NNLO results obtained with a cutoff of $\Lambda=0.9$ GeV and the blue dashed lines denote the relativistic NLO  results obtained with a smaller cutoff of $\Lambda=0.6$ GeV. The corresponding bands represent the uncertainties for a DoB level of 68\%. For comparison, we also show the LO relativistic results (black dotted lines) obtained with a cutoff of $\Lambda=0.6$ GeV and the two sets of non-relativistic N$^3$LO results  NR-N$^3$LO-Idaho ($\Lambda=0.5$ GeV, green dash-dotted lines)~\cite{Entem:2003ft,Machleidt:2011zz}  and NR-N$^3$LO-EKM (cutoff $=0.9$ fm, magenta short-dotted lines)~\cite{Epelbaum:2014efa,Epelbaum:2014sza}. The black dots denote  the PWA93 phase shifts~\cite{Stoks:1993tb}. The shaded regions denote that those data are not fitted and the corresponding relativistic results are predictions.}
\label{fig:EX-uncertainties}
\end{figure*}

In Ref.~\cite{Wang:2021kos}, we showed that the higher partial waves which do not receive contact  contributions up to NNLO can be well described  with a cutoff of 0.9 GeV. In addition, only the $^3D_3$ partial wave is sensitive to the cutoff, which implies that higher order LECs are needed for this particular partial wave to achieve renormalization group invariance. As a result, in the fitting of the LECs at NLO and NNLO, we fix the cutoff at 0.9 GeV. 

Following the strategy adopted in non-relativistic studies, e.g., Refs.~\cite{Epelbaum:1999dj,Epelbaum:2014sza}, we perform a global fit to the $np$ phase shifts for all the partial waves with total angular momentum $J\leq 2$~\cite{Stoks:1993tb}.~\footnote{For a justification of direct fits to phase shifts, see, Ref.~\cite{Escalante:2020nbq}. An alternative fit to the results of the Granada partial wave analysis~\cite{NavarroPerez:2013usk,NavarroPerez:2014npc,NavarroPerez:2016eli} is given in the Supplementary Material.} For each partial wave, we choose eight data points with laboratory kinetic energy $T_{\rm{lab}}=1,5,10,25,50,100,150,200$ MeV for the fitting. The $\chi^2$-like function to be minimized, $\tilde{\chi}^2$, is defined as 
\begin{equation}
    \tilde{\chi}^2=\sum_i(\delta^i-\delta^i_{\rm{PWA93}})^2,
\end{equation}
where $\delta^i$ are theoretical phase shifts or mixing angles, and $\delta^i_\mathrm{PWA93}$ are their empirical PWA93 counterparts~\cite{Stoks:1993tb}. A few remarks are in order. First, the $\tilde{\chi}^2$ defined above does not have proper statistic meaning, as no  uncertainties are assigned and the number of data fitted is a bit arbitrary (eight for each partial wave in the present study). Second, as the same uncertainties for the phase shifts and mixing angle are assumed, this necessarily put more weights on those partial waves of large magnitude, for example, $^1S_0$ and $^3S_1$. 

The so-obtained fitting results are shown in Fig.~\ref{fig:EX-uncertainties}, where the theoretical uncertainties are obtained via the Bayesian model explained in the Supplementary Material for a DoB level of 68\%.  The corresponding LECs are given in Table~\ref{Tab:LECs}. For comparison, we also show the non-relativistic N$^3$LO results obtained with different strategies for regularizing chiral potentials from Refs.~\cite{Entem:2003ft,Machleidt:2011zz} and Refs.~\cite{Epelbaum:2014efa,Epelbaum:2014sza} which are denoted as NR-N$^3$LO-Idaho and  NR-N$^3$LO-EKM, respectively.  Comparing the LO, NLO, and NNLO results as well as the uncertainties, it is clear that the chiral results are converging reasonably well. In addition, the LECs look quite natural, particularly, those at NNLO, as the magnitude of most of them is in between $1\sim10$, though $O_1$, $O_{14}$, $O_{15}$, and $O_{16}$ are perhaps a little bit large, but not abnormally large.

First we notice that the NLO and NNLO relativistic results describe the $np$ phase shifts very well up to $T_\mathrm{lab}=200$ MeV, at a level similar to the non-relativistic N$^3$LO results. Particularly interesting is that the NLO and the NNLO results also agree well with each other for $T_\mathrm{lab}\le 200$ MeV, while the NNLO results are in better agreement the PWA93 data for larger kinetic energies. This demonstrates that the chiral series converge well. On the other hand, for $^3F_2$, the NLO results are better, which can be attributed to  the compromise that one has to make to fit all the $J=2$ partial waves with five LECs to balance the large contributions of subleading TPE. It can be largely improved once the correlation between the $D$-waves with $J=2$ and $^3P_2$-$^3F_2$ are removed, i.e., the $D$-waves and $^3P_2$-$^3F_2$ are fitted separately or the cutoff is slightly modified. We note that in obtaining the  NR-N$^3$LO-Idaho results, the phase shifts of this channel were lowered by a careful fine-tuning of $c_2$ and $c_4$~\cite{Entem:2003ft}.

The $\tilde{\chi}^2$'s for each partial wave are given in Table~\ref{Tab:chi}. Judging from the total $\chi^2$, the quality of the relativistic fits is compatible to the non-relativistic N$^3$LO results. Comparing the NR-N$^3$LO-EKM results with the relativistic NNLO results, we find that although the total $\tilde{\chi}^2$'s are similar, they originate from different partial waves. The largest contribution to the total $\tilde{\chi}^2$ of NR-N$^3$LO-EKM comes from the $^1S_0$ partial wave while that of our NNLO results originates from the $^3D_2$ partial wave. It should also be noted that if we set the cutoff at 0.8 GeV, we can achieve a total $\tilde{\chi}^2$ 
as small as 5.3 (see the Supplementary Material for details), which is even smaller than that of NR-N$^3$LO-Idaho, which is about 9. However, as shown in Ref.~\cite{Wang:2021kos}, the $^3D_3$ partial wave cannot be well described with a cutoff of 0.8 GeV. Therefore, in the present work, we stick to the cutoff of 0.9 GeV.~\footnote{It should be mentioned that  the relatively large cutoff together with the pion-nucleon couplings $c_i$ determined from pion-nucleon scattering (which result in strongly attractive subleading TPE contributions in some channels) could lead to the appearance of deeply bound states (see Ref.~\cite{Reinert:2017usi} for a detailed discussion). Although they are beyond the range of the EFT  and  are irrelevant for the low-energy physics~\cite{Epelbaum:1999dj,Nogga:2005hy}, they could potentially complicate  many-body calculations (See Ref.~\cite{Nogga:2005hy} for a  solution).  We leave a study of their impact on relativistic ab initio calculations to a future work.}

To summarize, we constructed a relativistic chiral nucleon-nucleon interaction up to the next-to-next-to-leading order in covariant baryon chiral perturbation theory. The 19 low energy constants were fixed by fitting to all the partial wave phase shifts with total angular momentum $J\le2$. We obtained a good description of the PWA93 phase shifts. The next-to-leading order and the next-to-next-to-leading order results agree well with each other for $T_\mathrm{lab}\le 200$ MeV, while at higher energies the NNLO results agree better with the PWA93 phase shifts. This demonstrated the  convergence of the covariant chiral expansions. Given the quality already achieved in describing the $np$ phase shifts, the NNLO relativistic chiral $NN$ interaction  provides the much wanted inputs for relativistic ab initio nuclear structure and reaction studies. In particular, it may provide new insights into many long-standing problems, e.g., the $A_y$ puzzle, in combination with the  leading order relativistic $3N$ chiral force explored in Ref.~\cite{Girlanda:2018xrw}, which appears at NNLO.

This work is supported in part by the National Natural Science Foundation of China under Grants No.11735003, No.11975041,  and No.11961141004. Jun-Xu Lu acknowledges support from the National Natural Science Foundation of China under Grants No.12105006. Jie Meng acknowledges support from the National Science Foundation of China (NSFC) under Grants No. 11935003 and 
the National Key R \& D Program of China (Contracts No. 2017YFE0116700). Peter Ring acknowledges partial support from the Deutsche Forschungsgemeinschaft (DFG, German Research Foundation) under Germany’s Excellence Strategy—EXC-2094—390783311.
\bibliography{refs}

\newpage
\pagenumbering{gobble}
\begin{figure*}[t]
\centering
\includegraphics[page=1,width=1.0\textwidth]{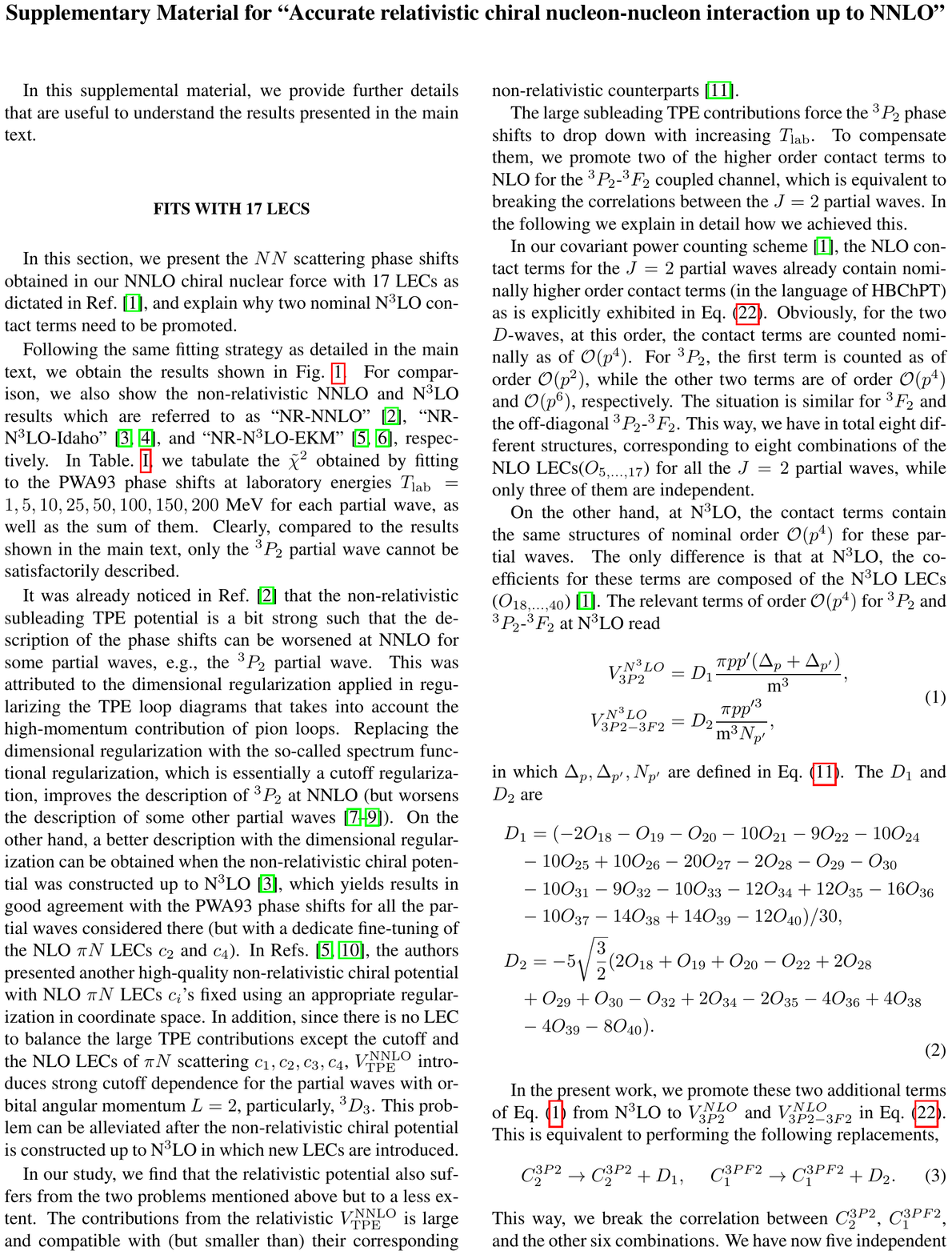}
\end{figure*}

\begin{figure*}[t]
\centering
\includegraphics[page=2, width=1.0\textwidth]{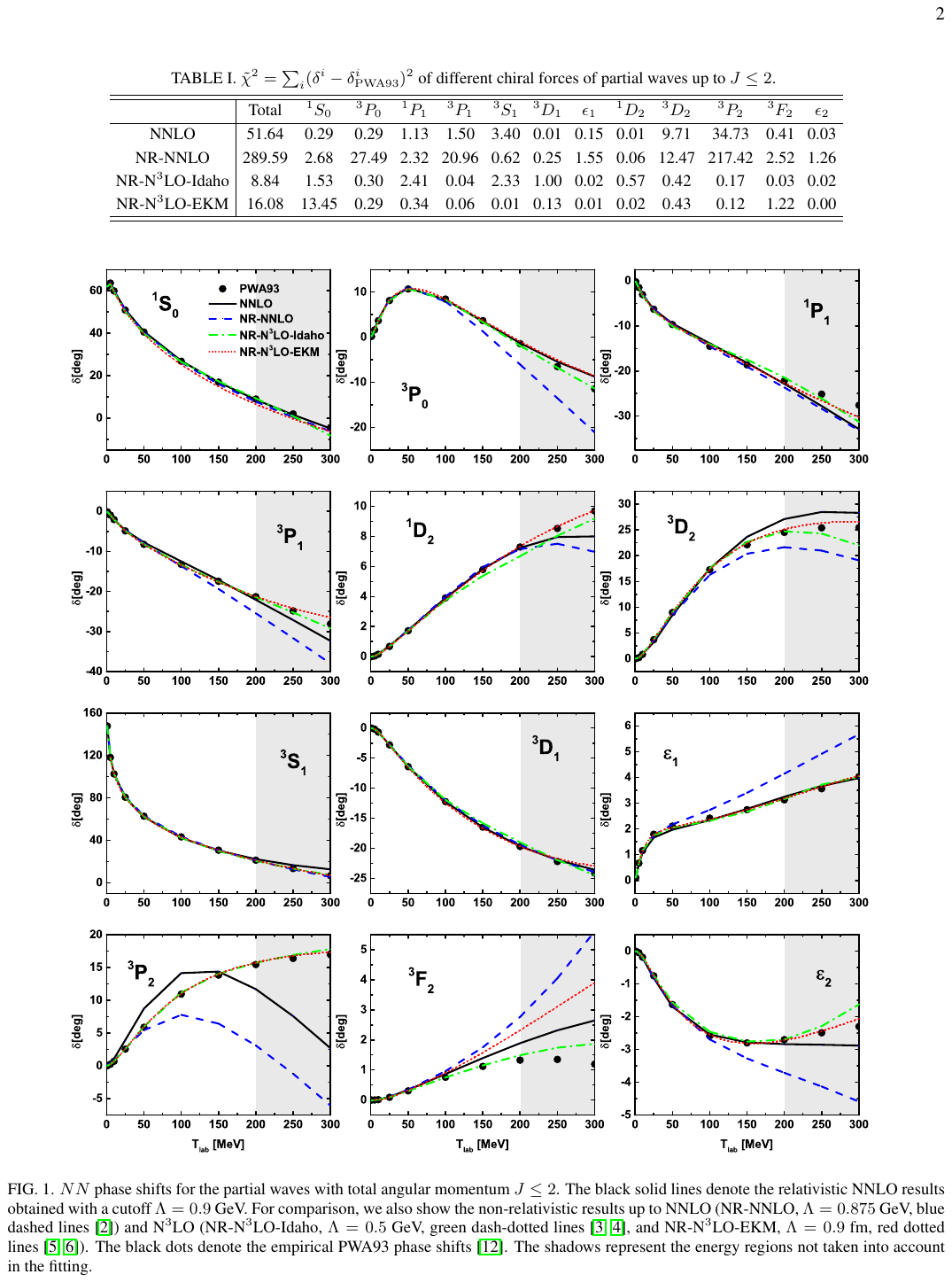}
\end{figure*}
\begin{figure*}[t]
\centering
\includegraphics[page=3, width=1.0\textwidth]{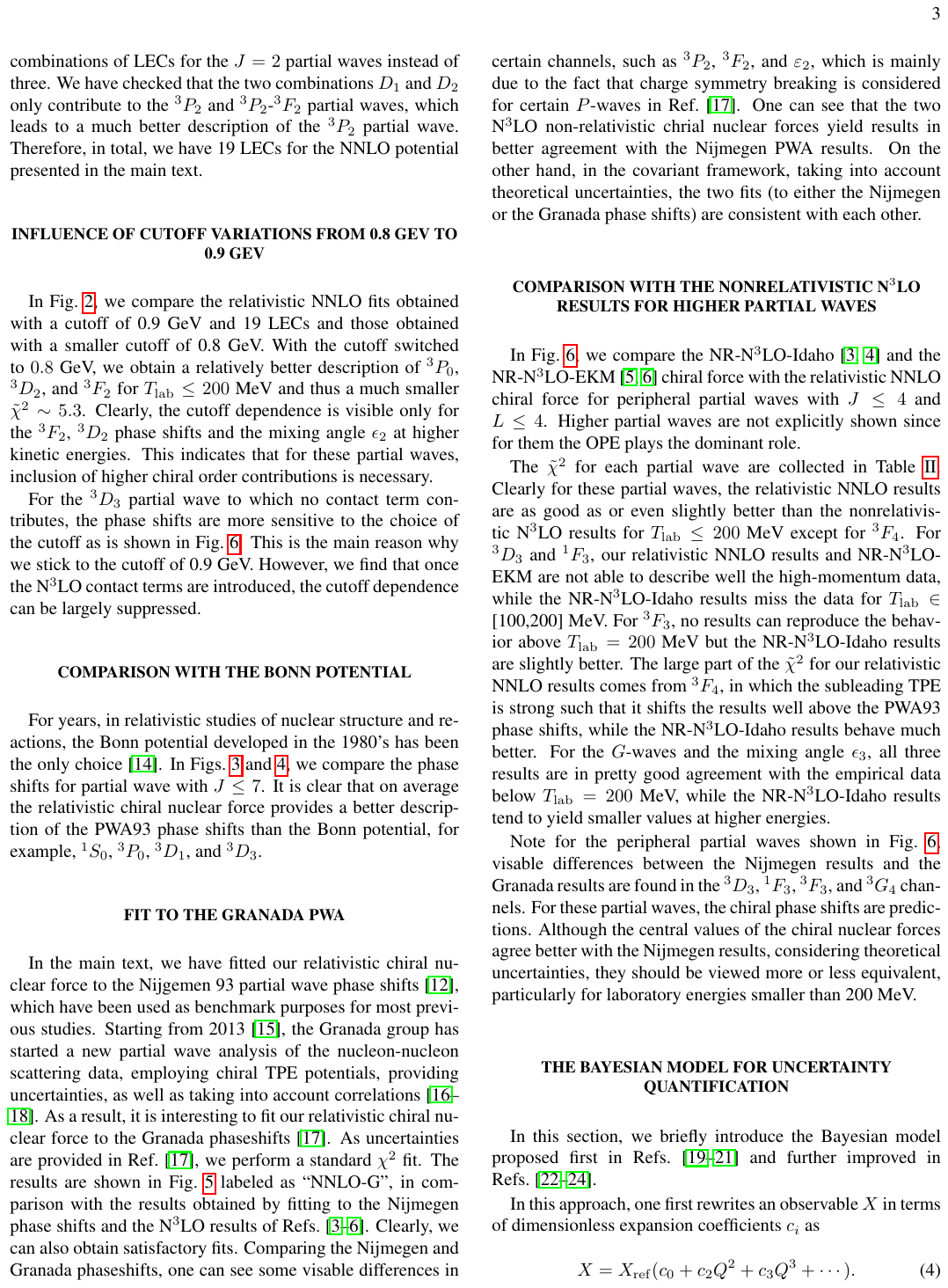}
\end{figure*}
\begin{figure*}[t]
\centering
\includegraphics[page=4, width=1.0\textwidth]{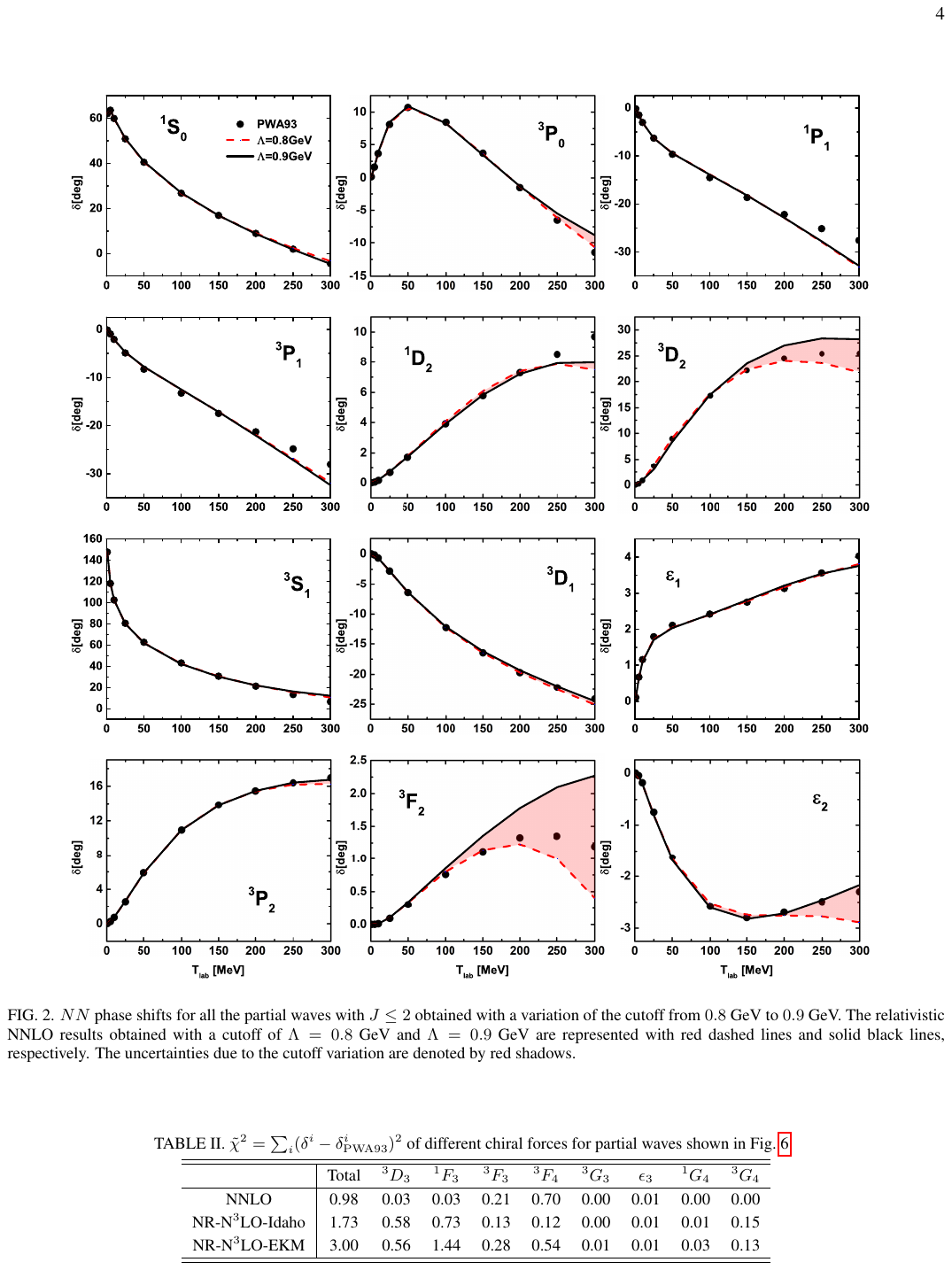}
\end{figure*}
\begin{figure*}[t]
\centering
\includegraphics[page=5, width=1.0\textwidth]{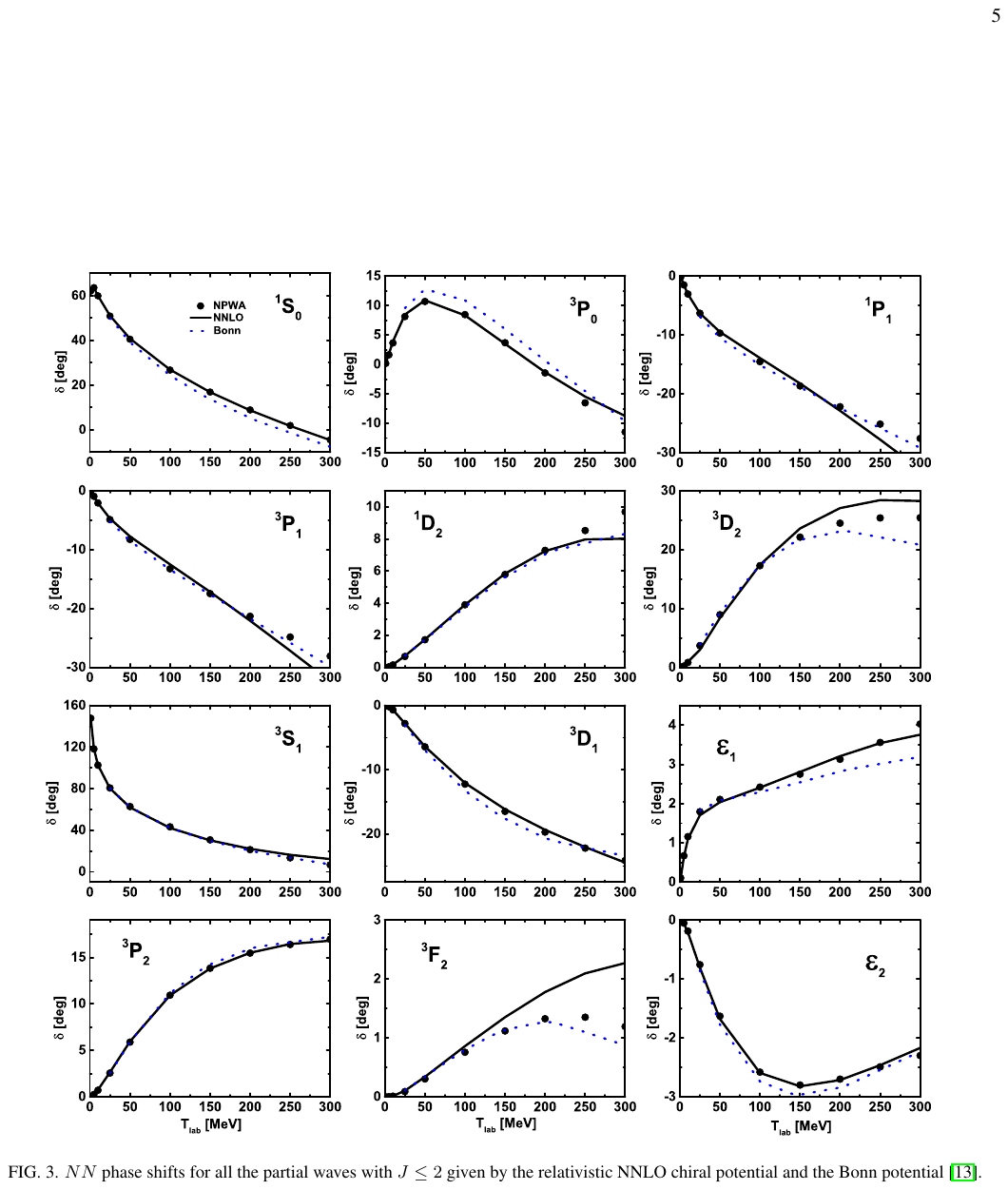}
\end{figure*}
\begin{figure*}[t]
\centering
\includegraphics[page=6, width=1.0\textwidth]{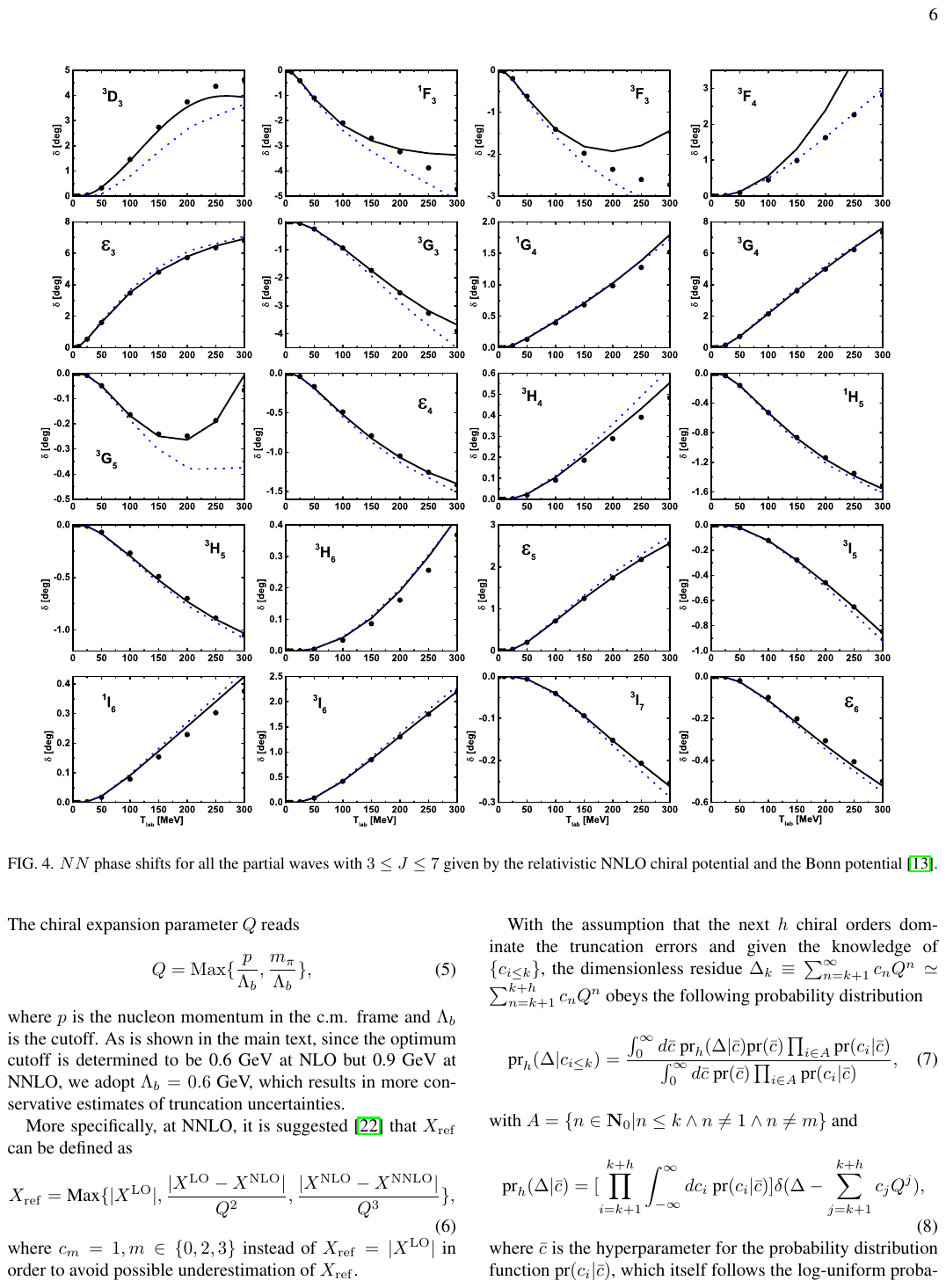}
\end{figure*}
\begin{figure*}[t]
\centering
\includegraphics[page=7, width=1.0\textwidth]{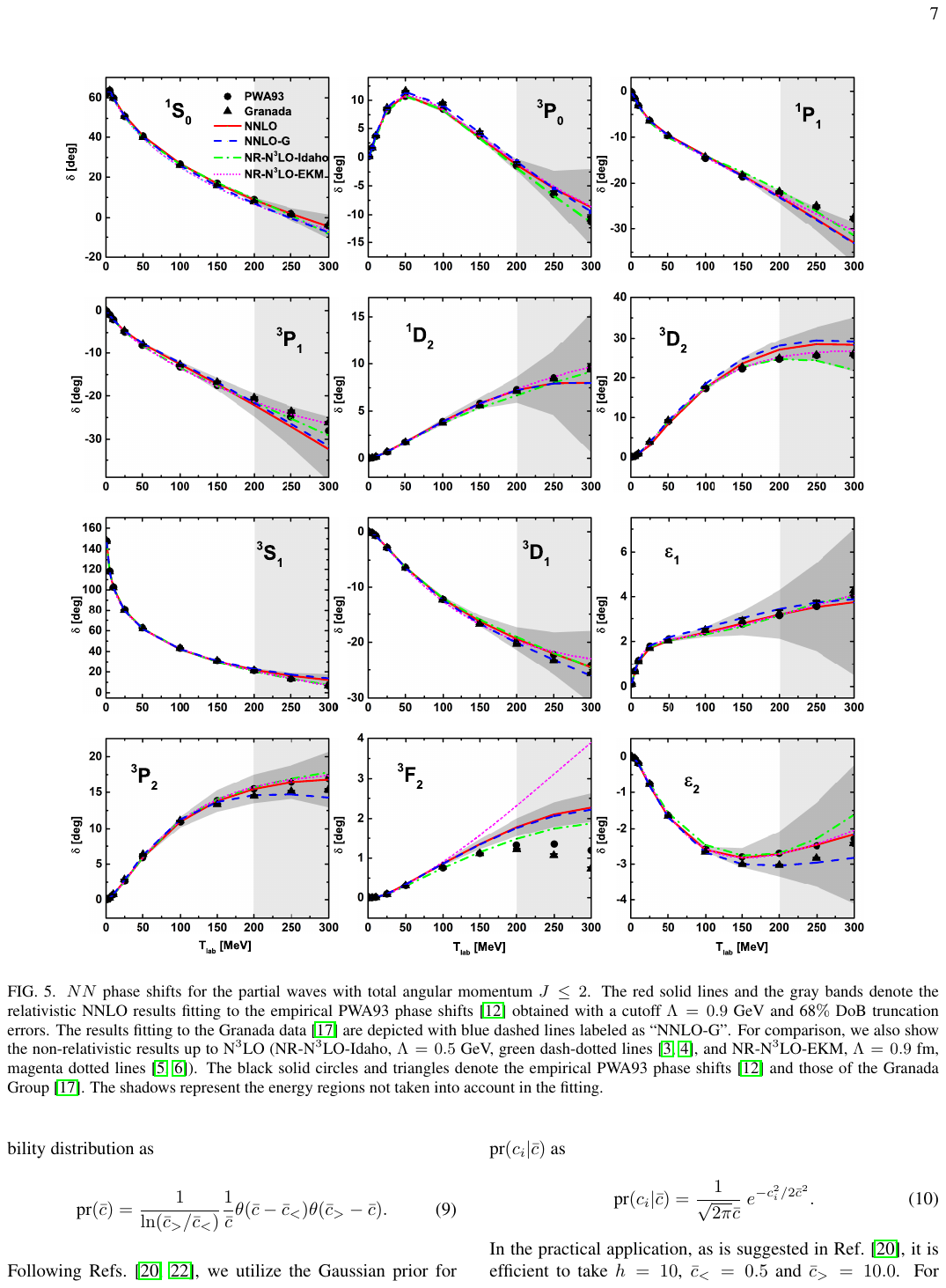}
\end{figure*}
\begin{figure*}[t]
\centering
\includegraphics[page=8, width=1.0\textwidth]{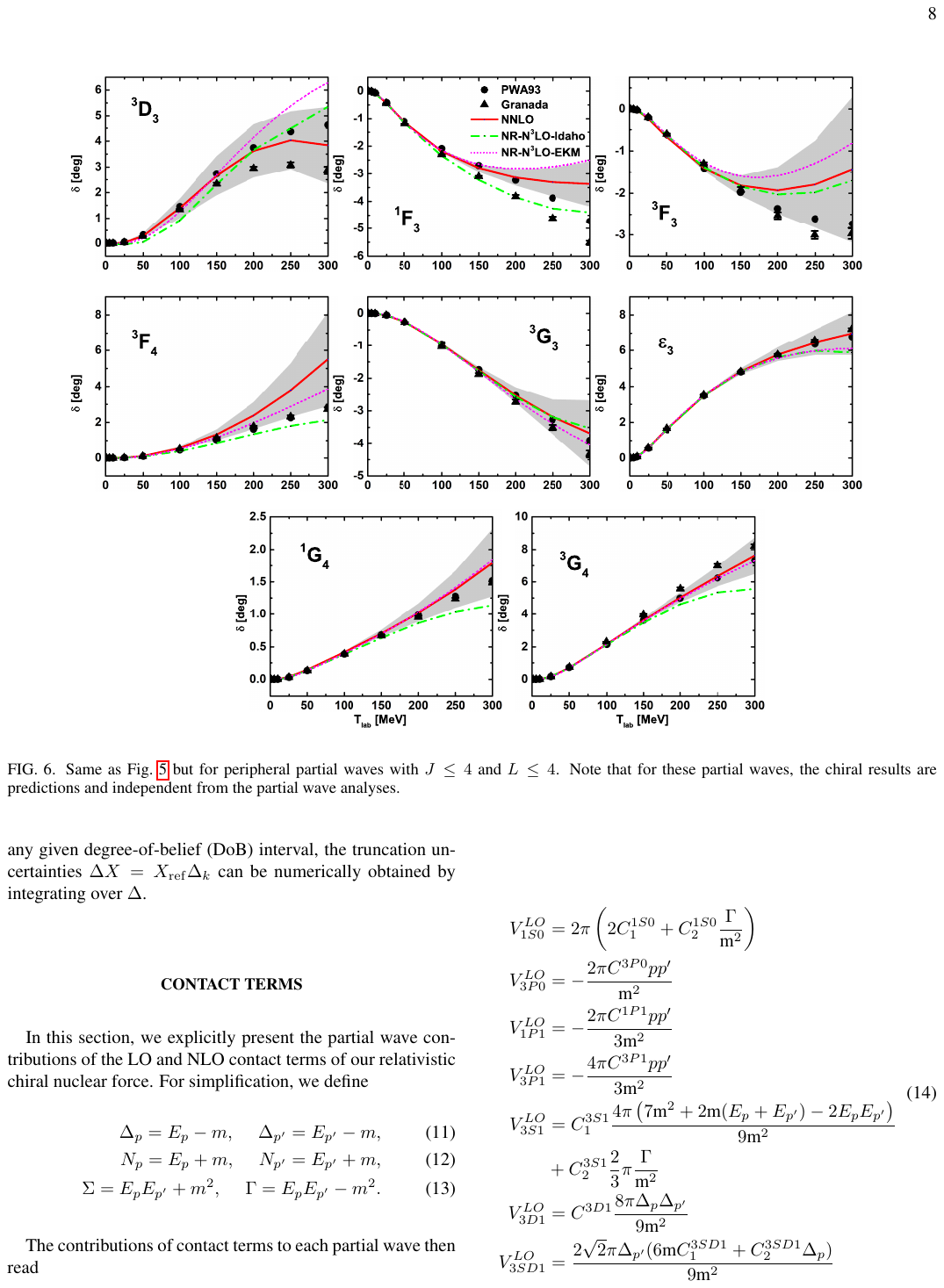}
\end{figure*}

\begin{figure*}[t]
\centering
\includegraphics[page=9, width=1.0\textwidth]{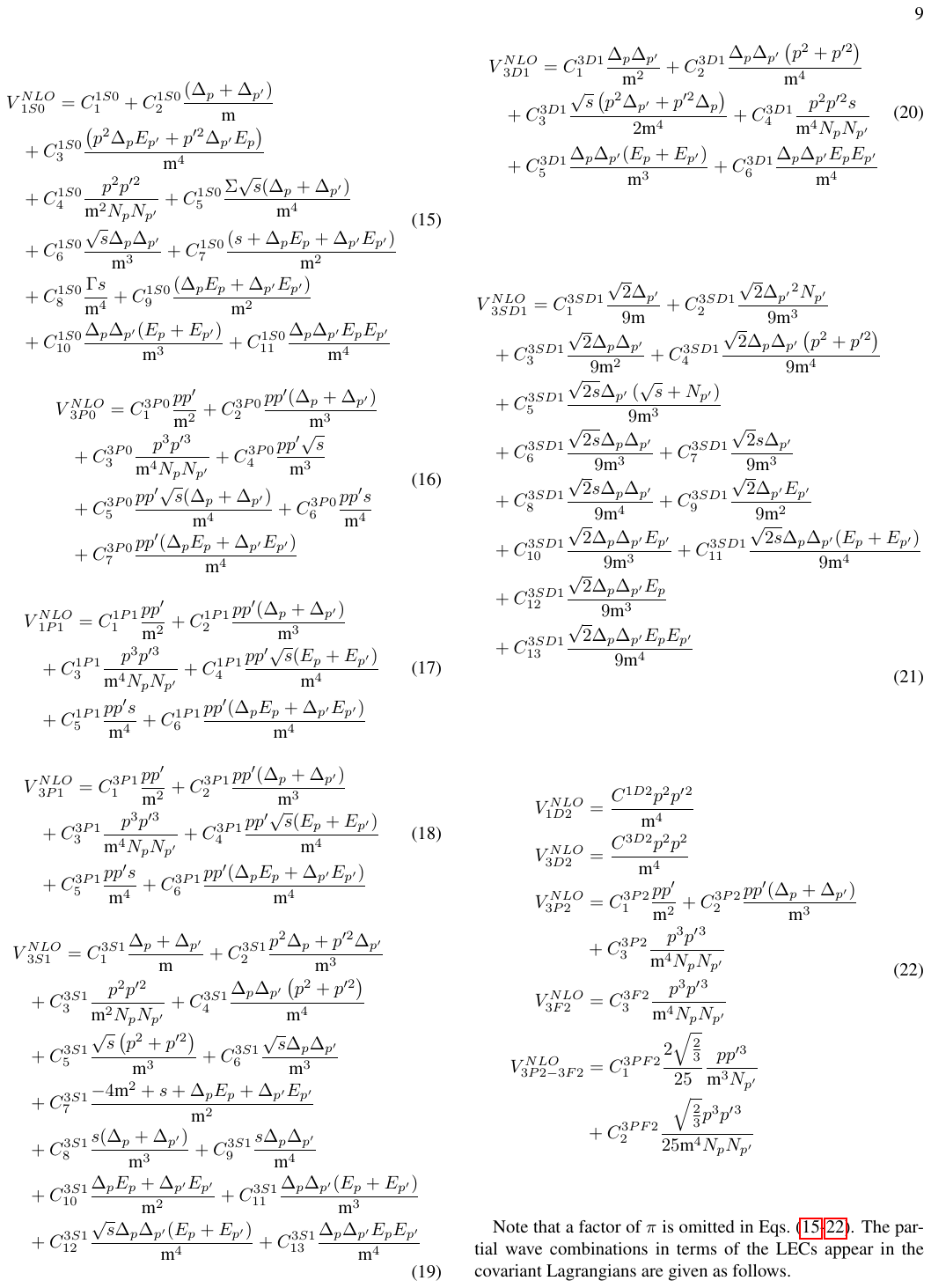}
\end{figure*}

\begin{figure*}[t]
\centering
\includegraphics[page=10, width=1.0\textwidth]{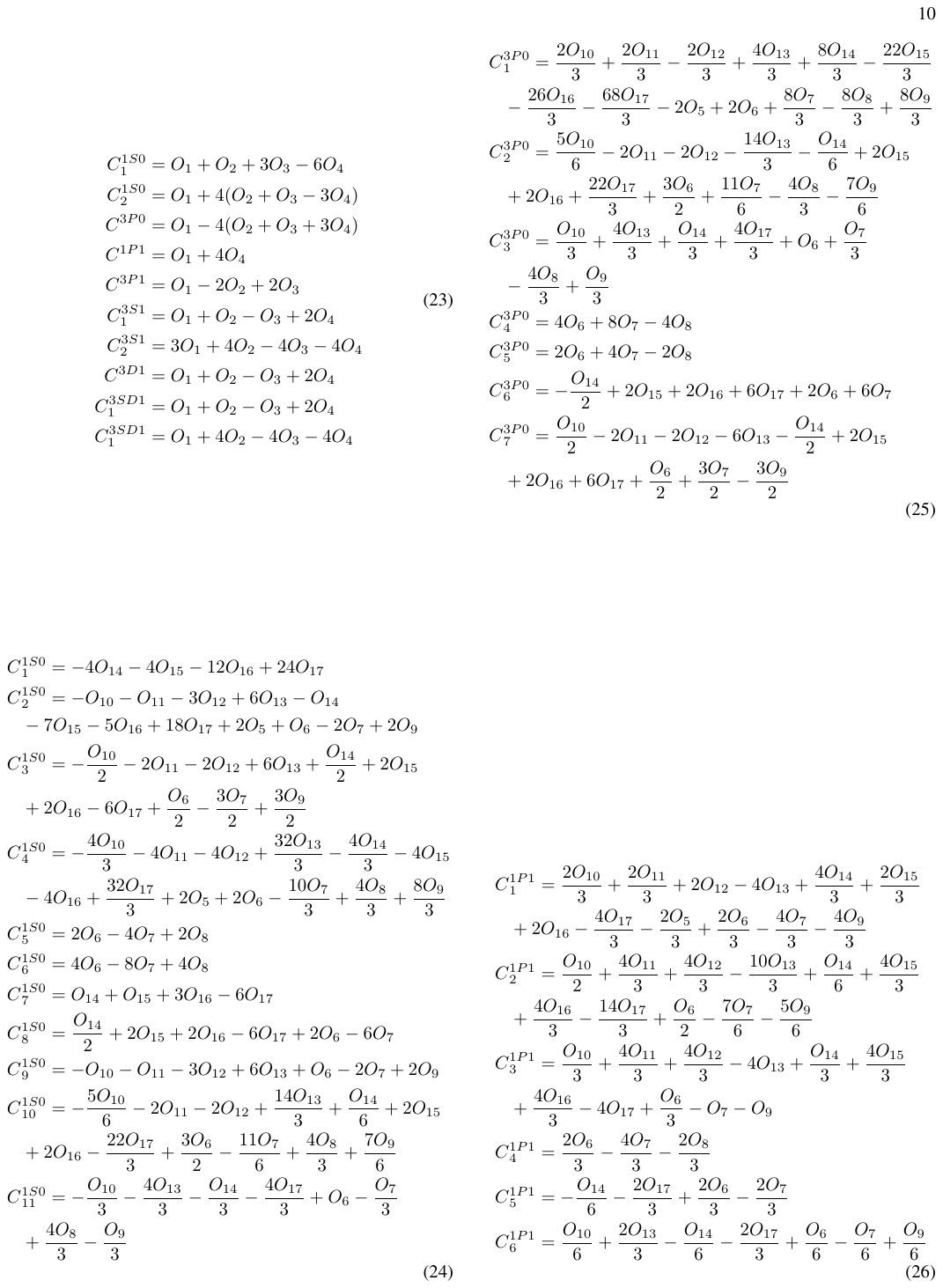}
\end{figure*}

\begin{figure*}[t]
\centering
\includegraphics[page=11, width=1.0\textwidth]{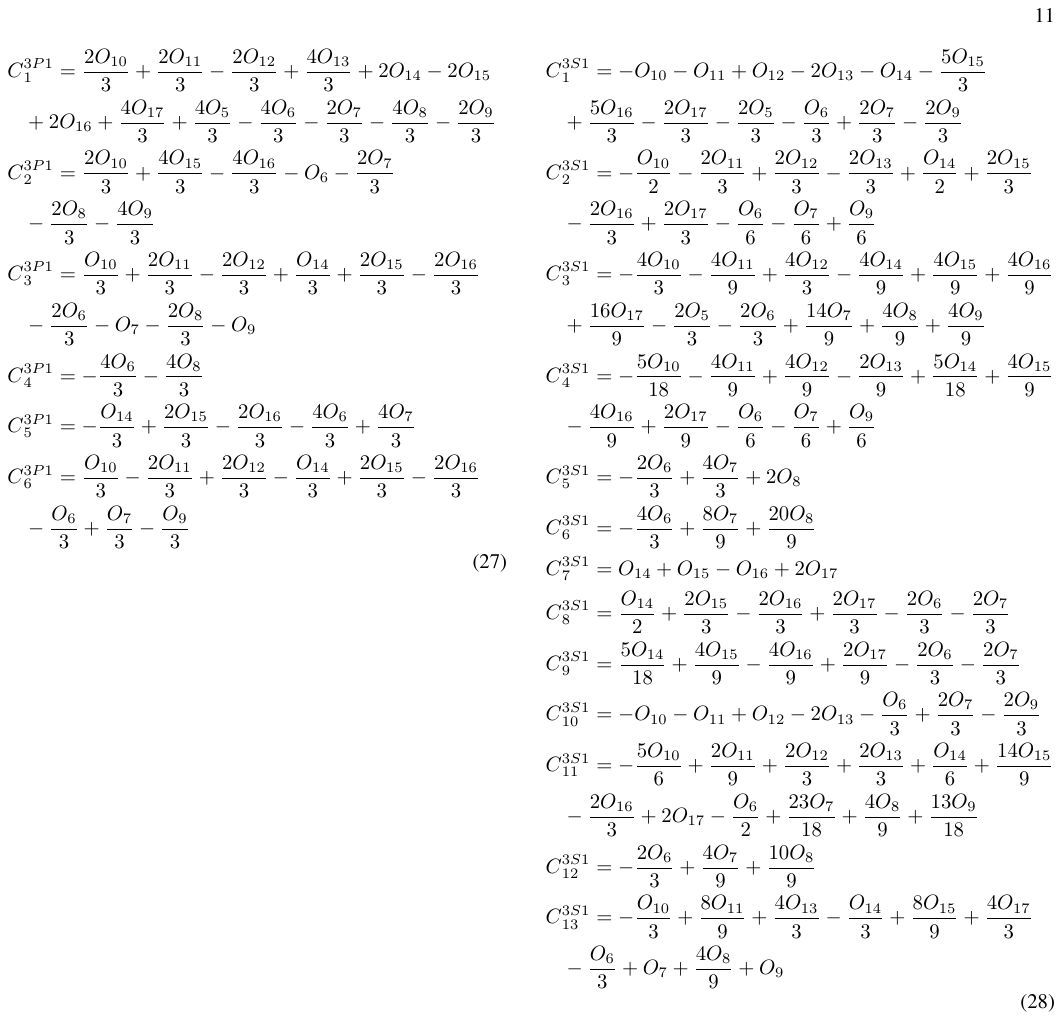}
\end{figure*}

\begin{figure*}[t]
\centering
\includegraphics[page=12, width=1.0\textwidth]{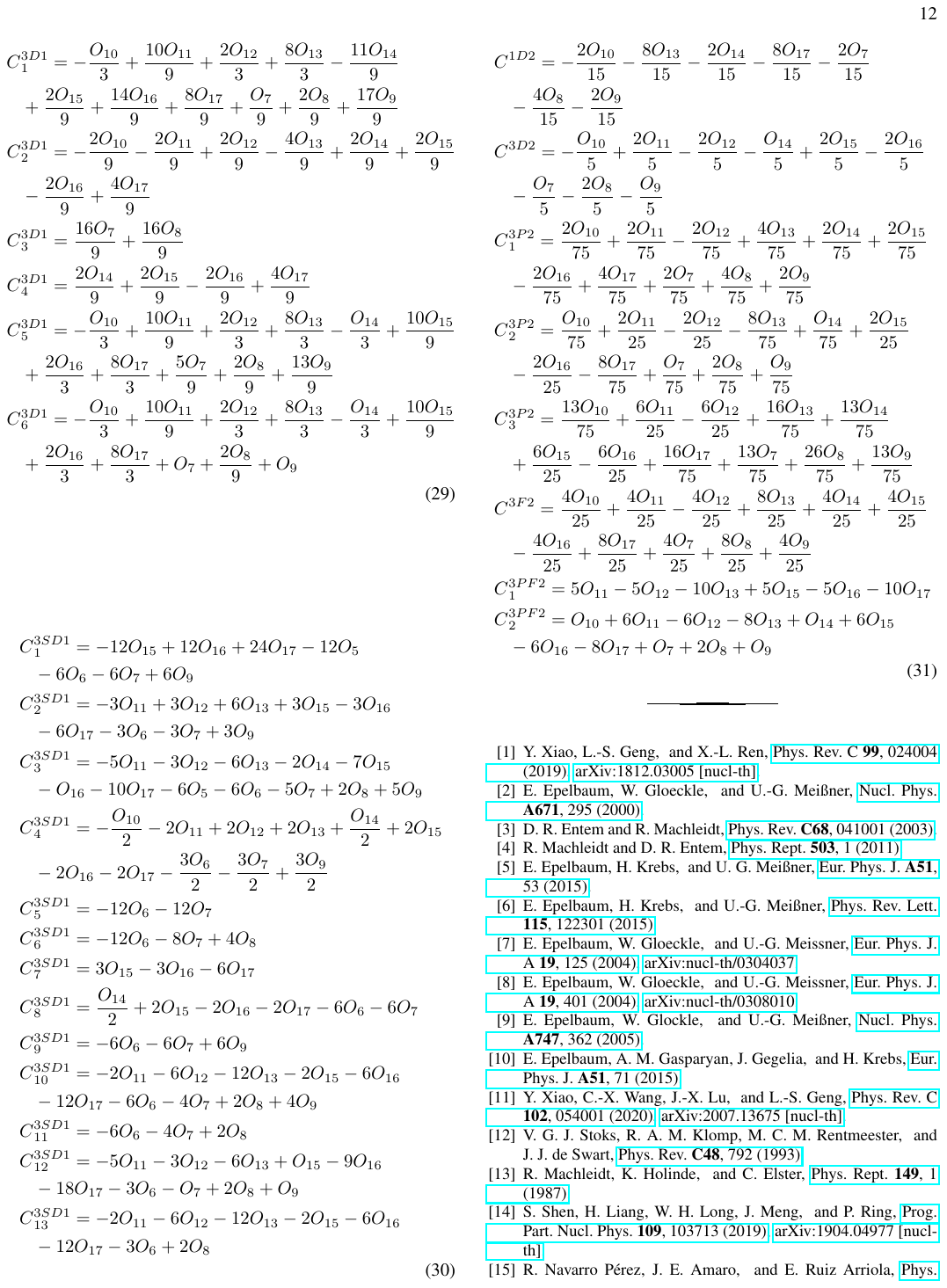}
\end{figure*}

\begin{figure*}[t]
\centering
\includegraphics[page=13, width=1.0\textwidth]{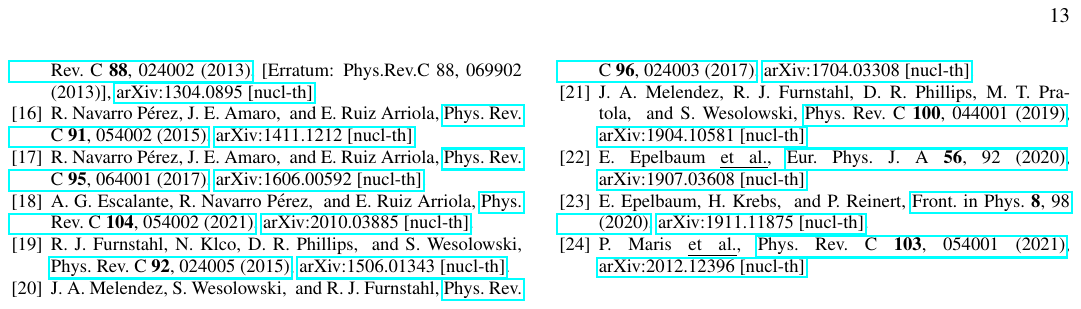}
\end{figure*}

\end{document}